\documentclass[twocolumn,notitlepage]{aastex62}
\pdfoutput=1 %for arXiv submission
\usepackage{amsmath,amstext,amsbsy,gensymb,chngpage}
\usepackage{booktabs,multirow}
\usepackage[T1]{fontenc}
\usepackage{apjfonts}
\usepackage[figure,figure*]{hypcap}
\usepackage{graphicx}

\newcommand{\B}[1]{\boldsymbol{#1}}
\newcommand{\U}[2]{$U(#1,#2)$}
\newcommand{\G}[2]{$G(#1,#2)$}
\newcommand{\rbold}[1]{#1}
\newcommand{\mbold}[1]{#1}
\newcommand{\rboldb}[1]{\rbold{#1}}

\defcitealias{hosek2015arches}{H15}
\renewcommand{\S}{Section } % We want to replace all section symbols with word "Section"

\shorttitle{Quintuplet Structure}
\shortauthors{Rui et al.}

\begin{document}

\title{The Quintuplet Cluster: Extended Structure and Tidal Radius}

\author[0000-0002-1884-3992]{\rbold{Nicholas Z. Rui}}
\affiliation{Department of Astronomy, University of California at Berkeley, CA, USA 94720}
%\email{nrui@berkeley.edu}

\author[0000-0003-2874-1196]{\rbold{Matthew W. Hosek Jr.}}
\affiliation{Department of Physics and Astronomy, University of California, Los Angeles, CA 90095, USA}
\correspondingauthor{Matthew W. Hosek Jr.}
\email{mwhosek@astro.ucla.edu}

\author[0000-0001-9611-0009]{Jessica R. Lu}
\affiliation{Department of Astronomy, University of California at Berkeley, CA, USA 94720}
%\email{jlu.astro@berkeley.edu}

\author[0000-0002-2577-8885]{William I. Clarkson}
\affiliation{Department of Natural Sciences, University of Michigan-Dearborn, 4901 Evergreen Road, Dearborn, MI 48128, USA}

\author[0000-0003-2861-3995]{\rbold{Jay Anderson}}
\affiliation{Space Telescope Science Institute, Baltimore, MD 21218, USA}
%\email{jayander@stsci.edu}

\author[0000-0002-6753-2066]{Mark R. Morris}
\affiliation{Department of Physics and Astronomy, University of California, Los Angeles, CA 90095, USA}
%\email{morris@astro.ucla.edu}

\author[0000-0003-3230-5055]{\rbold{Andrea M. Ghez}}
\affiliation{Department of Physics and Astronomy, University of California, Los Angeles, CA 90095, USA}
%\email{ghez@astro.ucla.edu}

\begin{abstract}
The Quintuplet star cluster is one of only three known young ($<10$ Myr) massive (M $>10^4$ M$_\odot$) clusters within $\sim100$ pc of the Galactic Center. In order to explore star cluster formation and evolution in this extreme environment, we analyze the Quintuplet's dynamical structure. Using the HST WFC3-IR instrument, we take astrometric and photometric observations of the Quintuplet covering a $120''\times120''$ field-of-view, which is $19$ times larger than those of previous proper motion studies of the Quintuplet. We generate a catalog of the Quintuplet region with multi-band, near-infrared photometry, proper motions, and cluster membership probabilities for $10,543$ stars. We present the radial density profile of $715$ candidate Quintuplet cluster members with $M\gtrsim4.7$ M$_\odot$ out to $3.2$ pc from the cluster center. A $3\sigma$ lower limit of $3$ pc is placed on the tidal radius, indicating the lack of a tidal truncation within this radius range. Only weak evidence for mass segregation is found, in contrast to the strong mass segregation found in the Arches cluster, a second and slightly younger massive cluster near the Galactic Center. It is possible that tidal stripping hampers a mass segregation signature, though we find no evidence of spatial asymmetry. Assuming that the Arches and Quintuplet formed with comparable extent, our measurement of the Quintuplet's comparatively large core radius of $0.62^{+0.10}_{-0.10}$ pc provides strong empirical evidence that young massive clusters in the Galactic Center dissolve on a several Myr timescale.
\end{abstract}

\keywords{astrometry -- Galaxy: center -- open clusters and associations: individual (Quintuplet) -- stars: kinematics and dynamics}

\section{Introduction}
The Quintuplet is a $\sim4.8$ Myr \citep{schneider2013ages}, $10^4$ M$_\odot$ \citep{figer1999hubble, figer1999massive} young massive cluster (YMC) near the Galactic Center (GC) at a projected distance of $\sim35$ pc from the supermassive black hole \citep{lang1999radio}.
The Quintuplet's close proximity to the GC makes it a benchmark for star formation in an extreme tidal environment, along with the Arches \citep[aged 2.5-3.5 Myr]{figer2002arches,schneider2013ages} and Young Nuclear \citep[aged 4-8 Myr]{lu2013ync} clusters.
It has long been speculated that the Arches and Quintuplet have followed similar formation and evolution scenarios.
As the older of the two clusters, the Quintuplet can be loosely regarded as a more evolved counterpart to the Arches.
These two clusters are thus representative cases of star cluster formation and evolution in the GC environment.

It is believed that most stars originate from star clusters \citep{lada1993environments}, but it is manifestly clear that a large population of stars exist outside of star clusters today.
Studies of star cluster structure and dynamics bridge this disconnect by providing a glimpse of the timescale of cluster dissolution, elucidating the introduction of stars into the general population.
Current observations of the Quintuplet allow us to directly observe the projected stellar density as a function of projected distance to the cluster center---the radial profile---which ultimately encodes information about a cluster's spatial extent and dynamical history.
\citet{portegies2010young} note that, counterintuitively, YMCs tend to lack sharp tidal truncations, a conclusion corroborated in part by observations of extragalactic star clusters \citep{mackey_globular_a,mackey_globular_b,mackey_globular_c} as well as the Arches cluster \citep{hosek2015arches}.

By comparing the radial profile at different stellar mass ranges, one can also probe for mass segregation.
Mass segregation is the concentration of more massive stars at the cluster center, and is thought to be a combination of primordial---imprinted at formation---and dynamical mass segregation, the latter of which is thought to occur on times comparable to the cluster's relaxation timescale \rbold{\citep{larson1982mass,bonnell1998mass,bonnell2001ms,mcmillan2007dynamical}}.
Mass segregation has been widely observed in YMCs such as the Arches \cite[hereafter \citetalias{hosek2015arches}]{stolte2002massseg,hosek2015arches}, Westerlund 1 \citep{brandner2008wd1}, and NGC 3603 \citep{harayama2008ngc3603}, as well as the Quintuplet itself \citep{hussmann2012pdmf}.

The pursuit of answers to such questions requires the identification of star cluster members from a given field. One recent method utilizes proper motions, identifying cluster members as stars which co-move with the bulk cluster \cite[e.g.,][\citetalias{hosek2015arches}]{stolte2008proper,stolte2014orbital}. Though the Quintuplet has been subject to proper motion studies in the past, such studies have been strongly limited by small fields-of-view and lack of sub-$0.5$ mas yr$^{-1}$ proper motion precision \citep{stolte2014orbital}.

The case for constructing a proper motion membership catalog extends beyond explorations of the cluster's structure.
Previous studies of YMC initial mass functions (IMFs) have favored top-heavy IMFs in both the Arches and Quintuplet \citep{espinoza2009massive,hosek2019imf,hussmann2012pdmf}, although some of these claims have been disputed \citep{shin-arches,shin-quintuplet}.
A wider-field proper motion-based membership catalog also allows for the measurement of a more precise cluster bulk motion which, combined with age information from the stellar population, helps to reconstruct the birthplace of the Quintuplet.
The location of the Quintuplet's birth and its internal structure and dynamics have far-reaching implications for stellar and star cluster formation.
For example, the question of whether the Arches and Quintuplet clusters formed from collisions between molecular clouds on $x_1$ and $x_2$ orbits in the Galactic bar \citep{stolte2008proper,stolte2014orbital}, from the tidal compression of molecular clouds during closest passage to Sgr A* \citep{longmore2013ssc,kruijssen2014clouds}, or some other mechanism is yet to be answered.

The study of the Quintuplet cluster is, however, not without its challenges.
The foremost hurdle is in distinguishing cluster members from the field population.
In this work, we identify stars having proper motions similar to that of the cluster proper as being gravitationally bound to the cluster.
Due to the relative independence of this approach from photometry, proper motion membership is fairly robust against biases from extreme differential reddening and contamination from the large field population.
This is in contrast to purely photometric means of cluster membership determination, such as using the isochrone to identify members.

Differential reddening on its own also presents a significant obstacle which needs to be characterized for accurate mass-flux conversion.
The Quintuplet cluster lies behind $A_V \sim 25$ magnitudes of visual extinction from dust in the foreground Galactic plane; thus it is only possible to study this cluster at infrared wavelengths.
Furthermore, extinction varies across the field, smearing out the color-magnitude diagram (CMD).

Presented in this study is an extended proper-motion catalog of the Quintuplet covering a field-of-view 19 times larger than that of any previous proper-motion catalog.
We produce the first proper motion-based radial profile of the Quintuplet cluster and assess the possibility and extent of mass segregation and asymmetric structure.
We discuss these results noting similar revelations regarding the Quintuplet's younger cousin, the Arches cluster, and other star clusters in the Milky Way disk.

\begin{figure} \label{3colorimg}
\includegraphics[scale=0.45]{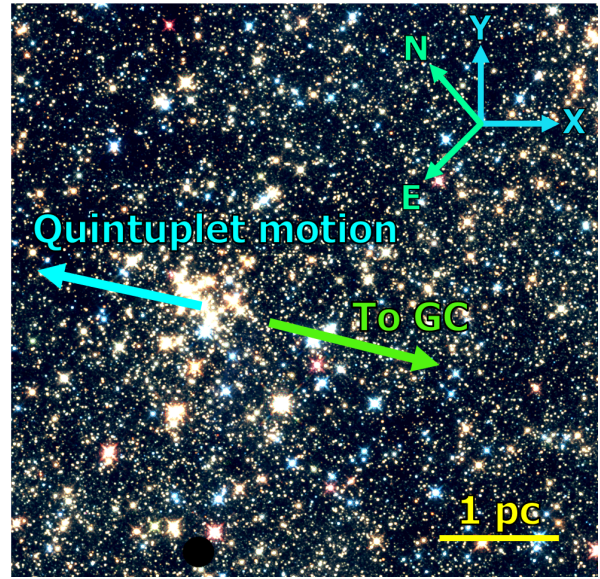}
\centering
\caption{A three-color image of the Quintuplet cluster constructed from HST WFC3-IR data \rbold{of our field of view}, with F153M in red, F139M in green, and F127M in blue. The black circular region at the bottom of the image is a known instrumental defect. We calculate membership probabilities for 10543 stars with proper motion errors $>$ 0.65 mas yr$^{-1}$ and an F153M magnitude $>$ 0.06 mag.}
\end{figure}

\section{Observations}

The Quintuplet cluster was observed with the \textit{Hubble Space Telescope} (HST) WFC3-IR camera four times over a six year baseline (from 2010 to 2016) using the F153M filter (1.53 $\mu$m) as well as the F127M and F139M filters in 2010 (1.27 $\mu$m and 1.39 $\mu$m, respectively) in programs HST-GO-11671, HST-GO-12667, HST-GO-12318 (PI: A. Ghez) and HST-GO-14613 (PI: J. Lu).
The images were downloaded from the Mikulski Archive for Space Telescopes\footnote{https://archive.stsci.edu/} and were processed using the standard HST pipeline, which performs flat-fielding and bias-subtraction on the images.
The data were processed using the CALWF3 code version 3.1 (28 December 2012) for the 2010-2012 epochs and version 3.3 (28 January 2016) for the 2016 epoch.

The observations are centered around $\alpha(J2000)=17^{\textnormal{h}} 46^{\textnormal{m}} 13.90^{\textnormal{s}}$, $\delta(J2000)=-28\degree 49' 48.00''$ with a position angle of $-42.3\degree$. All observations were made concurrently with the Arches data referenced in \citetalias{hosek2015arches}.
The images cover a field of view of $120''\times120''$ with a plate scale of 0.122 arcsec pix$^{-1}$.
The plate scale is derived from a comparison of WFC3-IR and ACS-WFC observations of the globular cluster $\omega$ Cen \citep{omegacent}.
Sub-pixel dithering was executed in order to compensate for the under-sampled PSF in the WFC3-IR F153M filter.
The F153M exposures are dithered using a 21-point spiral dither pattern with a point spacing of $\sim0\farcs4$.
The dithers also sample the full-range of pixel phases, which is critical for extracting precise astrometry.
The F127M and F139M exposures were dithered using a 4-point box pattern with a point spacing of $\sim0\farcs6$.
The observations are summarized in Table \ref{tab:observations}.

The HST observations are not centered directly on the cluster but are rather offset from the center of the cluster by $\sim25''$, covering regions at large radii from the Quintuplet center while still maintaining a reasonable coverage of the inner cluster.
This allows us to probe for structural features at large radii, increasing our sensitivity to mass segregation, tidal tails, and tidal truncation.

\begin{table}\label{tab:observations}

\caption{Summary of HST Observations}\label{observations}
\begin{center}
\begin{tabular}{l c c c c}
\hline
\hline
Date & Filter & Images & $t_{\mathrm{img}}$ (s) & N$_{stars}$ \\
\hline
\rbold{2010 August 10} & F127M & 12 & 600 & 54,173 \\
\rbold{2010 August 10} & F139M & 10 & 350 & 54,284 \\
\rbold{2010 August 16} & F153M & 21 & 350 & 54,402 \\
\rbold{2011 September 9} & F153M & 21 & 350 & 63,581 \\
\rbold{2012 August 12} & F153M & 21 & 350 & 63,852 \\
\rbold{2016 October 22} & F153M & 21 & 350 & 55,692 \\
\hline
\end{tabular}
\end{center}
Summary of the observations of the Quintuplet cluster with HST WFC3-IR. Observations span a six year baseline and three different filters: F153M, F139M, and F127M. Exposure times are given per image. \hfill
\end{table}

We largely follow the methods of \citetalias{hosek2015arches}, summarized here for clarity.

\section{Data Analysis}

\subsection{Stellar Measurement Extraction}

% I'm not so clear on what quint reduce actually is, if it's custom software or just standard, also is there a paper to cite by Jay Anderson or something
In order to extract individual measurements of stars' positions and brightness from the raw images, we use a reduction routine developed by J. Anderson (\texttt{KS2}; \citealp{anderson2008acs}).
\texttt{KS2} identifies individual stars in multiple dithered exposures and outputs high-precision astrometric and photometric measurements.
\rbold{PSF fitting is conducted by interpolating over a $3\times3$ grid of PSF models derived for the WFC3-IR camera.
Source detections are performed iteratively on the stack of exposures, and identified stars are removed at each pass in order to find fainter stars.
Source detection requires that stars be sufficiently isolated ($\sim1.5$ pixels apart), significantly above the sky background in the stack, and have a shape well-described by the PSF.}
Our extraction technique is described in more detail in Appendix A of \citetalias{hosek2015arches} \rbold{as well as in \citet{anderson2008acs}}.
This extraction produces a starlist of stellar positions and fluxes along with their uncertainties.
We adopt the RMS error over the individual exposures for the photometric uncertainty and the error on the mean for the astrometric uncertainty.
As in H15, we adopt the error on the mean for the astrometric uncertainty and the RMS error for the photometric uncertainty.
\rbold{We find that both the Arches and Quintuplet data sets show that the observed photometric residuals have a Gaussian distribution with a width better described by the RMS error than the error on the mean (see Figure 16 in \citetalias{hosek2015arches}).
On the other hand, the astrometric residuals are captured with the error on the mean. The source of this difference between the astrometric and photometric uncertainties is currently not clear.}
The mean astrometric uncertainties for stars brighter than magnitude 17 in F153M are 1.45 and 1.26 mas for X and Y, respectively.
The mean photometric uncertainties for these stars are 0.013, 0.007, and 0.010 mag for F153M, F139M, and F127M, respectively.
The astrometric and photometric uncertainties are presented as a function of magnitude in Appendix \ref{summaryplots}.
These choices are found to better capture the uncertainties for the astrometry (which are used when calculating the proper motions) and photometry.
In total, we measure 54,000--64,000 stars in each epoch and filter (see Table \ref{tab:observations}).

% Probably don't need this histogram, but it's fun to have for now
%Figure \ref{obs_histogram} shows a histogram of stars for which stellar measurements are obtained. Notably, the lack of completeness towards the center of the cluster is apparent---in addition to crowding, particularly near the center of the cluster, diffraction spikes due to brighter stars drastically reduce the number of locally detected stars. This reinforces the importance of a completeness correction in the Quintuplet for any meaningful analysis of its structure.

% Still have to quantify the coverage at various radii
% Also, aren't our observations actually worse for areal coverage somehow? !!!!!!!!!!!!!!!!!!!!!!!!!!!!!!!!!!!!!!!!!

\subsection{Alignment and Proper Motions}
% Check the "by eye to prime the alignment" wording
% Also make sure you have the facts 100% right re: search radius and how it works
% "bivariate, twelve-parameter, second-order polynomial fit"
To calculate the proper motions for Quintuplet stars, it is important to align (i.e.\rbold{,} cross-match and transform) the epochs into a common reference frame.
To this end, we perform an initial transformation with a bivariate, twelve-parameter, second-order polynomial fit using the position measurements obtained from \texttt{KS2}.
Stars must lie within $0.5$ pixels ($\sim60$ mas) of each other in each epoch to be identified as the same star.
After this first alignment, we trim out stars that are not observed in at least three epochs in order to estimate proper motions uncertainties.
We then fit the $x$- and $y$-proper motions of observed stars via linear fits, weighting by astrometric errors.
For each star, the proper motion fits are conducted relative to one error-weighted average time, $t_0$, in order to minimize fit bias from high uncertainty points.
The proper motion error is calculated as the square root of the diagonal component of the covariance matrix corresponding to a given proper motion component and we ignore any cross-covariances. 

\begin{figure}[h!] \label{propermotionfillbetween}
\includegraphics[scale=0.4]{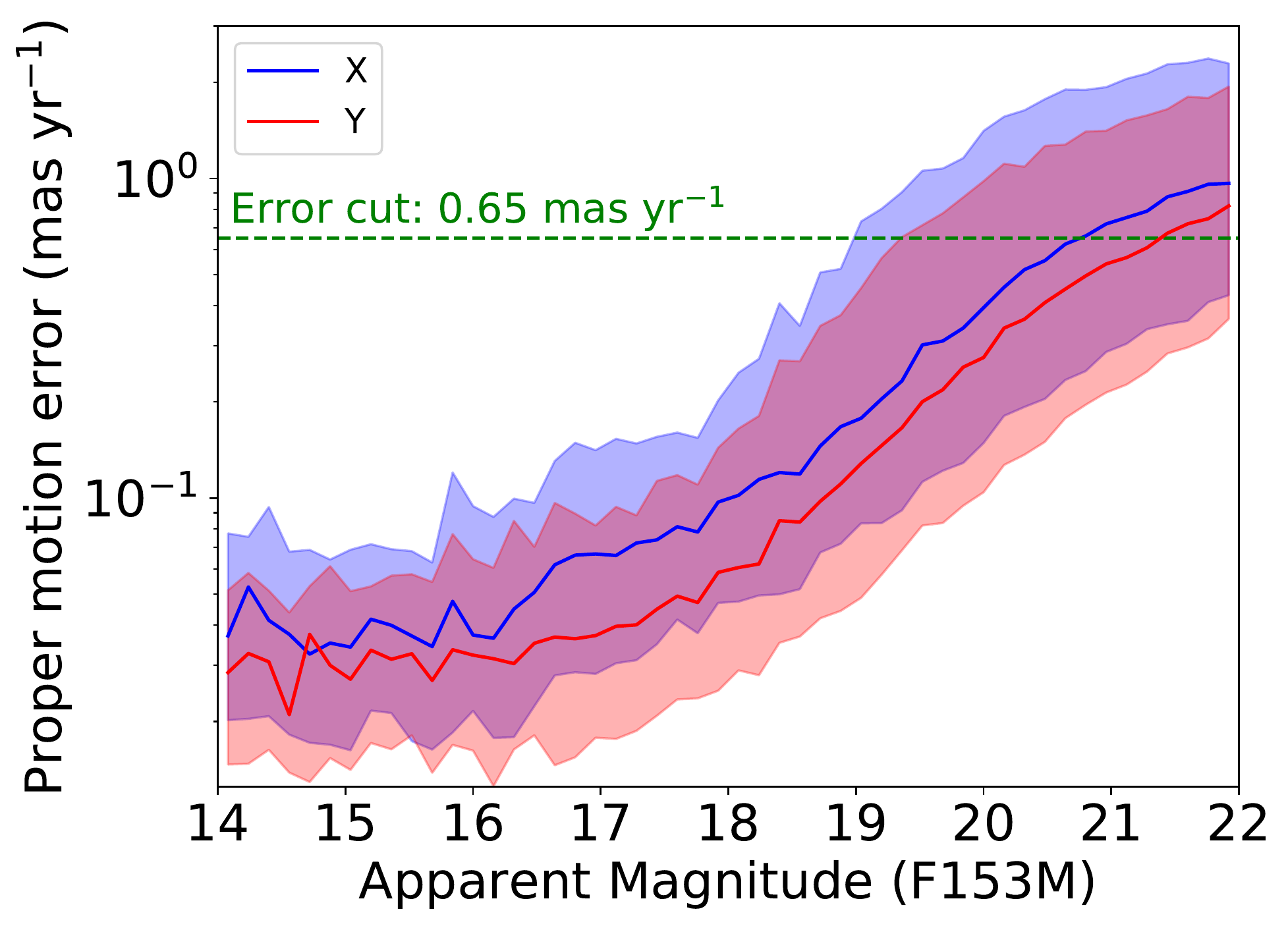}
\centering
\caption{Median $x$ and $y$ proper motion errors versus magnitude, with the shaded region being the $1\sigma$ regions. The green dashed line indicates the proper motion error ceiling of 0.65 mas yr$^{-1}$ applied for the Gaussian mixture model of the cluster/field kinematics (\S\ref{membership}).}
\end{figure}

However, the initial transformation could be biased by the fact that the reference stars used in the alignment have spatially-dependent proper motion structure.
To address this, we perform a preliminary 3-Gaussian mixture model to calculate initial proper motion membership probabilities (see \S\ref{membership}).
Using stars with kinematic membership probability $P^i_{\textnormal{PM,0}}>50\%$ (Section \ref{finalmembership}) as reference stars, we conduct a second alignment, again via second-order fit.
We stress that the membership probabilities used for realignment are not the membership probabilities used in the final analysis.
In total, 1,352 reference stars are used for the realignment. Overall, we calculate proper motions for 51,782 stars.
The proper motion errors as a function of F153M magnitude are shown in Figure \ref{propermotionfillbetween} and are typically $\sim0.04$ mas yr$^{-1}$ at the bright end. The proper motions are well-behaved along $y$ and show some systematic errors along $x$ based on analysis of the $\chi^2$ distribution (see Appendix \ref{summaryplots}). However, there is no trend between goodness-of-fit and magnitude or radius in the cluster, so the impact on our radial profile analysis is minimal.

\subsection{Image Completeness} \label{completeness}
Observational selection effects depending on magnitude and position are heavily influence the observed star counts.
We characterize this image completeness by generating an artificial star grid of 650,000 stars, planting them into the individual exposures, and attempting to recover them \rbold{one at a time} with \texttt{KS2} using the same methods as applied to the real data.
Of the 650,000 planted stars, 500,000 have fluxes which are randomly drawn from the Quintuplet cluster/field catalog.
Their magnitudes are perturbed by a Gaussian with a width corresponding to their photometric errors in each filter, with the perturbation being applied in magnitude space.
400,000 of these stars are planted uniformly on the images, and 100,000 are planted with a centrally-concentrated profile to increase sampling near the center of the cluster. The remaining 150,000 stars are generated in order to better capture the completeness at the bright and faint ends of the CMD, with 75,000 stars covering the faint end of the distribution and 75,000 stars covering the bright end of the distribution. For both the faint and bright star grids, 50,000 stars are planted uniformly over the image and 25,000 are planted with a centrally-concentrated profile.

In order to accurately mirror the selection effects on the real star sample, the artificial star population's astrometric and photometric errors must match those of the observed data.
As in \citetalias{hosek2015arches}, we find that our artificial stars lack the characteristic astrometric and photometric error floor of the observed population.
Thus, we manually enforce these error floors by summing the artificial star errors with the observed error floor in quadrature.
In \citetalias{hosek2015arches}, this was attributed to errors in the PSF\rbold{---}since artificial stars are generated with an assumed PSF, they are thus not expected to have such errors (see Appendix B in \citetalias{hosek2015arches} for some discussion of this).
In order to mitigate the impact of false detections of brighter stars in the vicinity of faint planted stars, we also reject retrieved artificial stars whose extracted positions deviate from the planted position by more than half a pixel or whose extracted magnitudes deviate from the planted magnitudes by more than 0.5 mag.

Proper motions and proper motion errors are then calculated for the artificial stars with detections in at least three epochs, mimicking the corresponding calculation for the observed stars. We find a slight discrepancy between the real and artificial proper motion errors, where the artificial errors are overestimated due to unknown systematics.
We correct the proper motion errors of the artificial stars by calculating the average proper motion error vs.~magnitude in both the real and artifical samples and subtracting the offset from the artificial stars.  
The corrected artificial proper motion distribution matches the empirical distribution.

% Note that the matchup mag cut is going to be replaced by a magnitude dependent cut most likely
In order to mimic the cuts applied to the real Quintuplet data, we tag as detections those stars satisfying the following conditions: (1) the star's position on the image be covered by at least 75\% of images available in that filter/epoch combination; (2) the star's magnitude error in that filter be no more than $0.06$ mag; (3) the star's proper motion error in either the $X$ or $Y$ direction across the image be no more than $0.65$ mas yr$^{-1}$; and (4) the star be recovered in at least three out of four F153M epochs.

\begin{figure}
\includegraphics[scale=0.35]{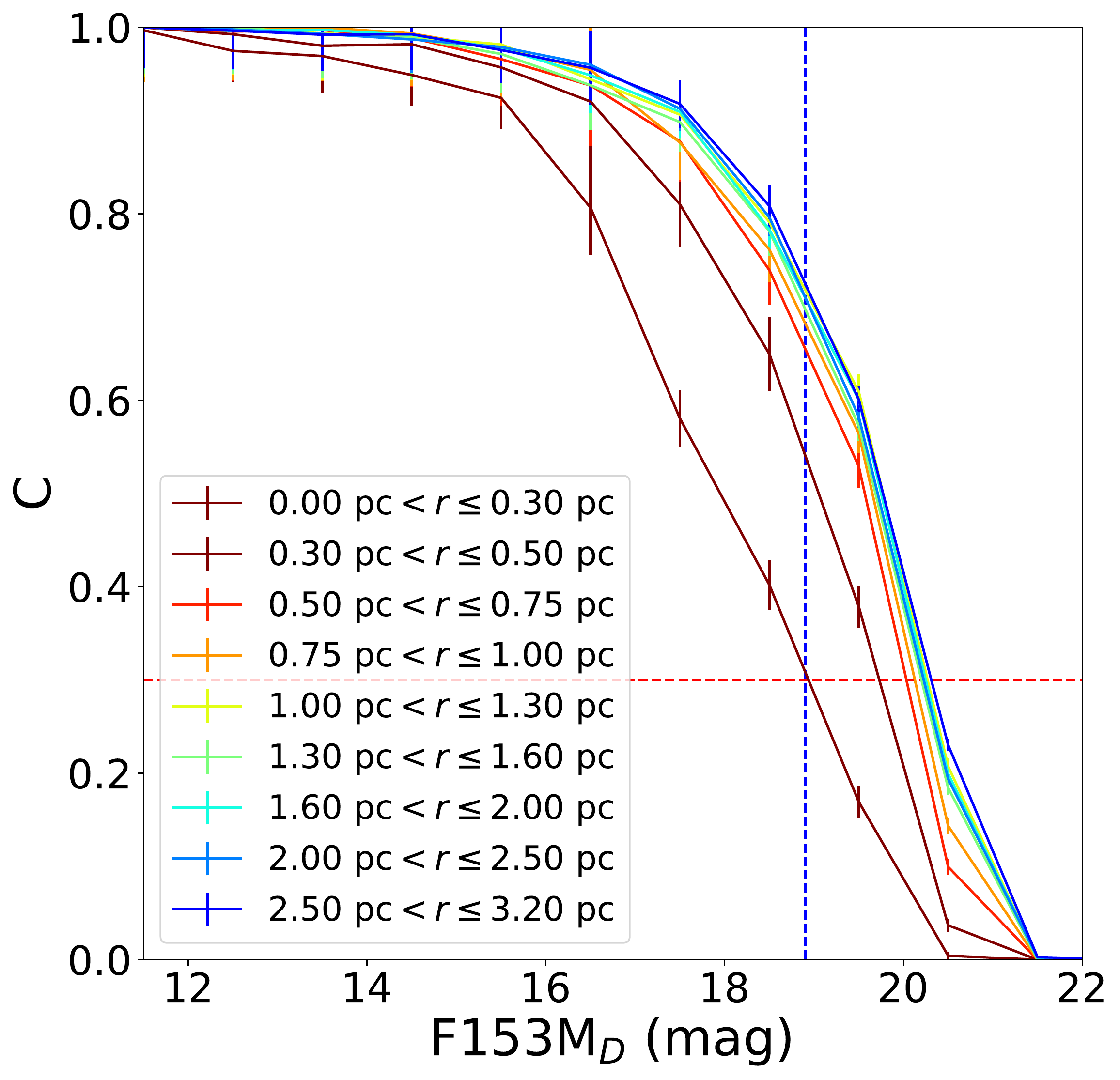}
\centering
\caption{\rbold{Image completeness fraction $C$ plotted against differentially dereddened F153M magnitude} for the Quintuplet field derived from star-planting simulations. Each line corresponds to a different radial bin. The red dashed line marks $30\%$ image completeness, and the blue dashed line indicates a differentially dereddened F153M magnitude of 18.9 mag. We adopt this as the magnitude limit for stars that can be included in the profile fit. \rbold{For the radial profile, image completeness corrections are assigned via differentially dereddened magnitudes so that the magnitude cut applied to our sample corresponds to a consistent stellar mass across the field.}} \label{completeness-curves}
\end{figure}

The artificial stars are binned as a function of \rboldb{differentially dereddened} F153M magnitude and radius.
\rboldb{Dereddened} magnitude is binned from 11 to 25 mag, with an equal spacing of 1 mag.
The radial bins are bounded by 0, 0.30, 0.50, 0.75, 1.00, 1.30, 1.60, 2.00, 2.50, and 3.20 pc, chosen to be roughly logarithmic, but with the inner bins increased in size in order to make sure there are a sufficient number of artificial stars used to calculate the image completeness in each bin.
The \rbold{image} completeness is interpolated with respect to differentially dereddened F153M magnitudes to generate an image completeness curve for each radial bin (\S\ref{extinction} describes the calculation of these differentially dereddened magnitudes).
These image completeness curves are shown in Figure \ref{completeness-curves}.
\rboldb{These curves are used to infer the existence of undetected stars when calculating the radial profile (see Section \ref{radialprofilefit}).
We find that interpolating by observed rather than differentially dereddened magnitudes makes almost no difference to the profile, with binned surface densities being recovered well within errors.}

\section{Extinction Mapping} \label{extinction}
The Quintuplet lies in a highly extinguished line of sight \citep{Cardelli:1989fp}, making an extinction correction crucial.
Because the amount of extinction varies across the field, the extinction must be mapped spatially.

The red clump (RC) is a population of red giant stars with relatively consistent intrinsic colors and luminosities \citep[e.g.,][]{stanek1994color,girardi2016RC}.
This makes RC stars invaluable as ``standard crayons''---examining the degree to which RC stars are extinguished as a function of position on the detector enables us to construct an extinction map of the Quintuplet field-of-view \citep[e.g.,][]{nataf2016interstellar}.
For this analysis, we assume that all extinction is due to foreground material and thereby disregard distance-dependent reddening effects.
This assumption performed adequately for the \citetalias{hosek2015arches} analysis of the Arches cluster and, when applied, significantly corrects the differential extinction in the Quintuplet, as discussed below.

\begin{figure*} \label{unsharp_mask}
\begin{center}
\includegraphics[scale=0.4]{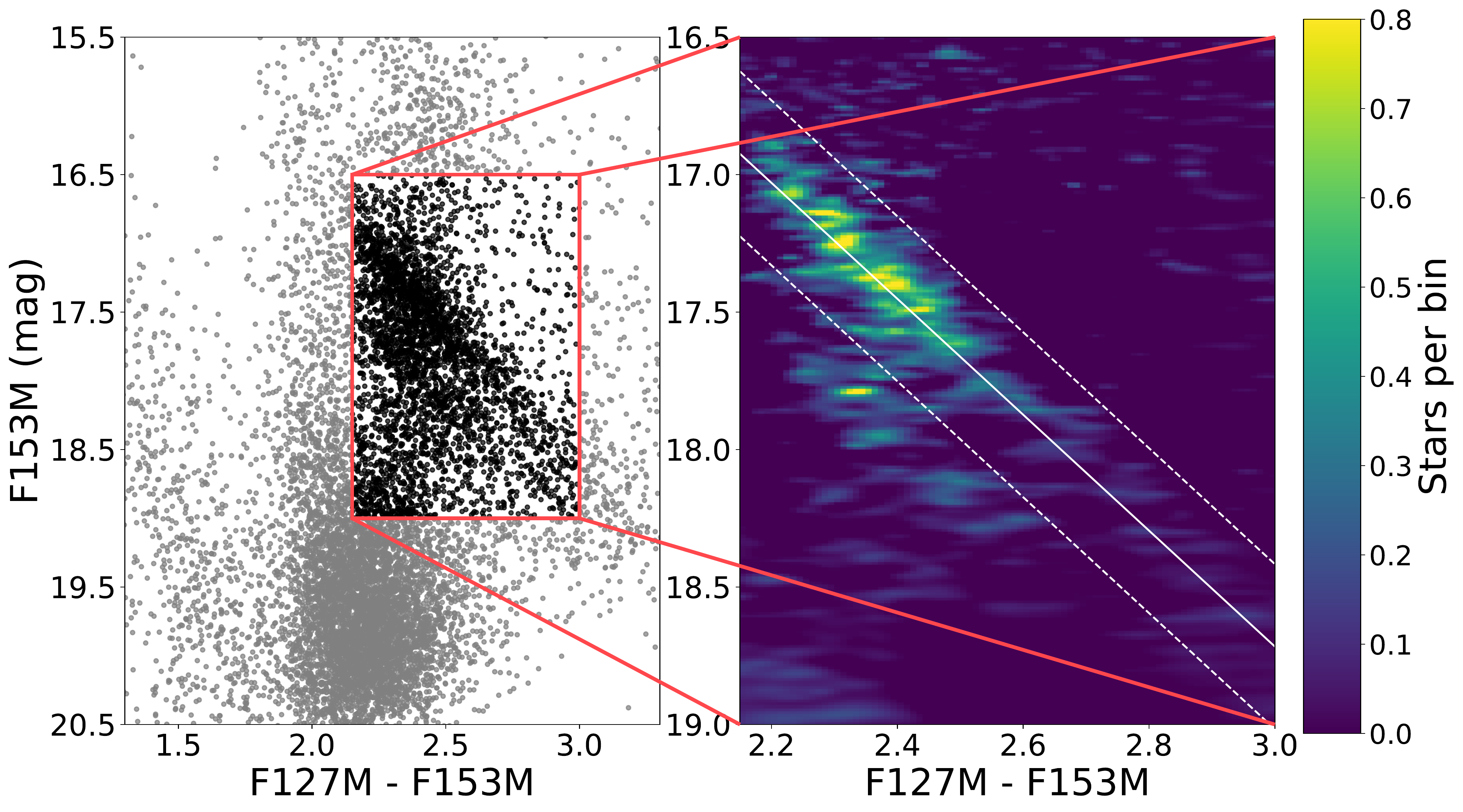}
\end{center}

\caption{\textit{Left}: The CMD of the red-clump (RC).
\textit{Right}: Hess diagram of the CMD after unsharp masking, where the RC sequence becomes significantly clearer. The region demarcated by the white dashed lines is the CMD selection region for RC stars used to construct the extinction map. The RC appears on the CMD (Figure \ref{CMDs}) as a smeared out stellar population in color.}
\end{figure*}

% "While the details of the method are summarized in \citet{de2015hubble} provide further details" (this line was removed)
Due to the large degree of differential reddening, the RC is realized as a smeared-out population on the CMD.
While the RC's presence is clear on the CMD, it is difficult to determine the RC's exact boundaries, required to maximize the number of RC stars used to generate the extinction map without accepting too many non-RC contaminant field stars.
Therefore, we apply the method of unsharp masking, which has been used to accentuate the boundaries of the RC on the CMD \citep[e.g.,][]{de2015hubble}.
Summarized briefly, unsharp masking involves the subtraction of a blurred version of a figure from the un-blurred version, suppressing low-frequency detail.
We perform unsharp masking on a Hess diagram \citep[][a 2D histogram in color-magnitude space]{hess1924probleme}, and use the result to perform an iterating linear fit which produces a very precise capture of the RC on the CMD (Figure \ref{unsharp_mask}).

% Return to this when you have a final cut on the CMD for the RC, you probably have this already but maybe if you toy around with spline maps or something later you'll get a better map
% Also, I've updated the discussion about the reddening vector, synthetic photometry, etc. is there a paper that I can cite for where the model is from? (if have time, can just check popstar myself)
% I would like to justify the use of the Hosek+17 law over Nishiyama+09, but I'm not familiar with the details, something about space telescope?
% Note, there is a mention of the B-spline interp here. If you change the map, change this paragraph too
We fit the unsharp-masked Hess diagram RC population with a line.
Stars with an F153M lying within 0.3 mag of this line with colors $2.15<$ F127M$-$F153M $<3.00$ are considered RC stars.
After making the requisite cuts on the CMD to obtain the RC, we remove stars with kinematic cluster membership probabilities greater than or equal to $0.05$ in order to minimize contamination from the cluster sequence, which overlaps with the RC primarily between the F127M$-$F153M color 2.15 to 2.5 (kinematic cluster membership assignment is described in \S\ref{finalmembership}).
We also enforce a magnitude error cut of $0.05$ mag on F153M, F139M, and F127M, in order to ensure accurate photometry for the extinction map.

We generate a reddening vector using an RC model from a 10 Gyr PARSEC stellar isochrone \citep{bressan_parsec} at solar metallicity, which has been found to describe the average bulge stellar population \citep{zoccali_agemetal,clarkson_bulge}.
Our model matches the effective temperature $T_{\textnormal{eff}}=4700$ K and a surface gravity $\log g=2.40$ cgs measured for solar-metallicity RC stars in the \textit{Hipparcos} catalog \citep{mishenina_clump}.
The RC is assumed to be at a distance of 8 kpc.
We apply a corrected version of the extinction law of \citet{hosek18extlaw}, which characterizes optical/near-infrared extinction in highly reddened regions and has been successfully applied to the study of the Arches initial mass function \citep{hosek2019imf}.
\citet{hosek18extlaw} report that variation in distances to RC stars causes a color spread $\mathrm{F127M}-\mathrm{F153M}\simeq0.14$, which corresponds to an uncertainty in the \rbold{A\textsubscript{Ks}} of $\simeq0.15$.

\begin{figure} \label{extinction_map}
\includegraphics[scale=0.45]{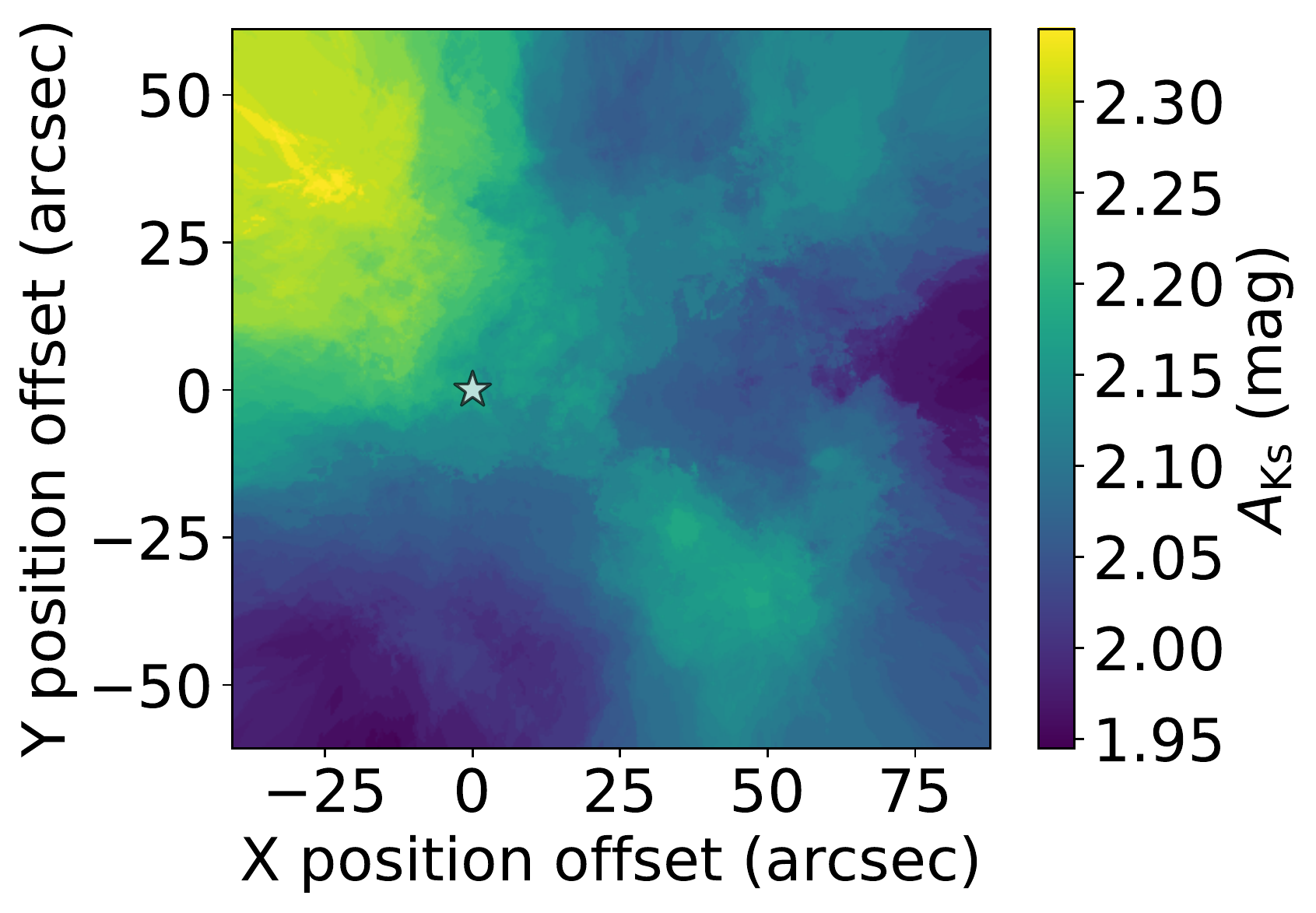}
\centering

\caption{The extinction map generated by analyzing the differential reddening of red clump stars in the field. This map is generated by assigning to a pixel the median $A_{\textnormal{Ks}}$ of the nearest $N=100$ RC stars. The white star represents the cluster center, taken to coincide with star Q12.}
\end{figure}

\begin{figure*} \label{CMDs}
\includegraphics[scale=0.45]{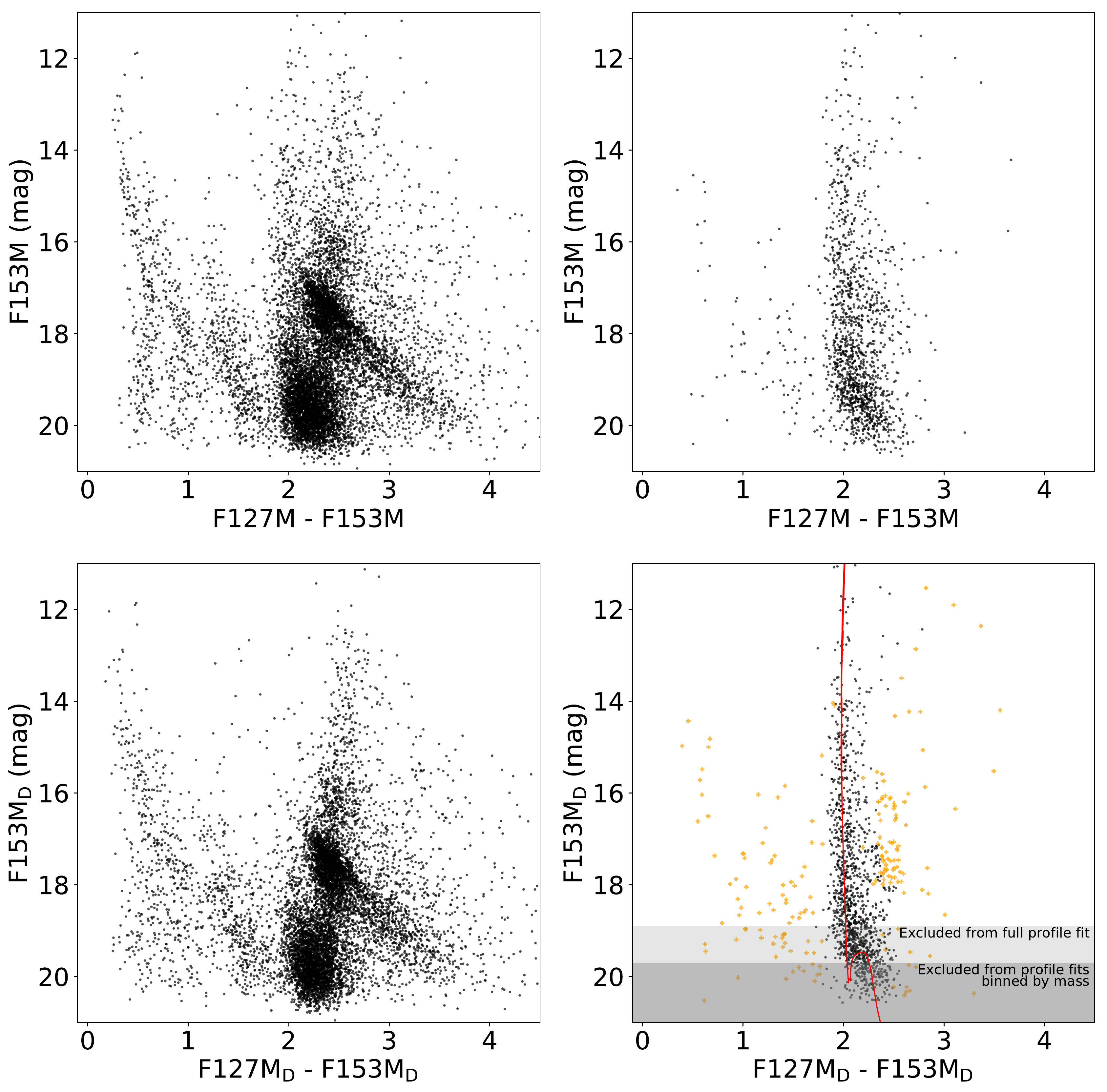}
\centering
\caption{\rbold{\textit{Top left}:} CMD of stars used in the Gaussian mixture model fit (stars with proper motion error $<0.65$ mas yr$^{-1}$ and F153M magnitude error $<0.06$ mag). Note that the point size on the leftmost panel is reduces to show substructure in the RC. 
\rbold{\textit{Top right}:} CMD of kinematically determined cluster members (stars \rbold{for which} $\mbold{P^i_{\textnormal{PM}}}>30\%$) before differential extinction correction. 
\rbold{\textit{Bottom left}: CMD of unlikely cluster members ($P^i_{\textnormal{PM}}\leq0.3$) using magnitudes na\"ively dereddened with our extinction map. Note that, since such stars generally lie in the cluster foreground or background, it is invalid to assume that the extinction map correctly characterizes their differential reddening.}
\rbold{\textit{Bottom right}:} CMD of kinematically determined cluster members after extinction correction. Magnitudes are dereddened to cluster members' median $A_{\textnormal{Ks}}$ value of 2.12 \rbold{mag}. Stars which are kinematically categorized as cluster members but excluded as color outliers are shown as orange crosses. The cluster sample appears to be consistent with a cluster isochrone assuming $A_{\textnormal{Ks}}=2.12$ \rbold{mag}, an age of 4.8 Myr, and a distance of 8 kpc to the cluster, shown in red. The isochrone was generated using Geneva evolution models with rotation \citep{Ekstrom2012} for main sequence/evolved stars and Pisa evolution models \citep{Tognelli2011} for pre-main sequence stars. Stars which are excluded from the full profile (\S\ref{radialprofilefit}, \S\ref{king-model}, \S\ref{elsonfit}) and mass binned profile fits (\S\ref{procedure-mass-seg}, \S\ref{masssegregation}) due to low \rbold{image} completeness are indicated by the gray shaded regions.}
\end{figure*}

The $A_{\textnormal{Ks}}$ values for the observed RC stars are calculated by matching the color of each star to its corresponding color on the reddening vector.
These values are used to create an extinction map with a nearest neighbor method where each pixel is assigned the median $A_{\textnormal{Ks}}$ value of its $N=100$ closest RC stars (Figure \ref{extinction_map}).
We perform 3$\sigma$ clipping during this step to reject outliers.
Stars in the catalog are then assigned the $A_{\textnormal{Ks}}$ value of their pixel location.
Stars' magnitudes are then dereddened to the mean extinction $A_{\textnormal{Ks}}=2.12$ mag of probable cluster members ($P^i_{\textnormal{PM}}>30\%$, Section \ref{finalmembership}).
Hereafter, a subscript D after a filter refers to the value of that magnitude after dereddening to this median extinction value (for example, F153M\textsubscript{D} refers to the differentially dereddened F153M magnitude).
The full sample CMD, along with the original and dereddened CMDs of likely cluster members, is shown in Figure \ref{CMDs}.

% Note that there is an RC 0.05 mag err cut on the RC selection

% Commented out, replace this with a better map if you ever get one
%We also try a nearest-neighbor method, where we assign each pixel the median extinction of its $N$ nearest neighbors for various values of $N$, rejecting RC stars with magnitude errors above $0.05$. However, we find that the B-spline interpolation map visibly reduces the spread around the isochrone of likely Quintuplet members, so we use that map for final $A_{\textnormal{Ks}}$ value assignments. It is likely that the B-spline's interpolation map's better ability to deredden the cluster to the isochrone is due to the B-spline's conformity to the continuous nature of the actual extinction across the field, whereas a nearest-neighbor method's discrete $A_{\textnormal{Ks}}$ assignment creates discontinuous patches with nonexistent sub-structure.

% Be sure about what's going on here, since I didn't write any of this code
% Be VERY SURE to check this over, because I'm writing this from memory
% Is this even linear? In what way (how many times) is it (does it) iterative(iterate)?

\section{Modeling the Cluster}

\subsection{Modeling Cluster and Field Kinematics}
\label{membership}
To identify cluster members, we assign membership probabilities based on proper motions. Stars that co-move with the cluster are more likely to be Quintuplet members, which are gravitationally bound to the cluster. We model the cluster and field kinematics with a 4-Gaussian mixture model in order to assess the cluster membership probability of a given star.

% Towards the end of this paragraph, this starts to get really similar to Matt's paper
With measurements of stellar coordinates in proper motion space for each star assumed to be independent of those of all other stars, we write the likelihood function $\mathcal{L}$ for the set of measured stars as
\begin{equation}
\mathcal{L}=\prod^N_{i=1}L_i
\end{equation}
\noindent where $L_i$ is the component of the likelihood function corresponding to the $i$th star and $N$ is the number of stars used in model-fitting. For each star, $L_i$ is defined as
\begin{equation}
L(\B{\mu}_i)=\mbold{\sum_{k=1}^K}\frac{\pi_k}{2\pi\lvert\B{\Sigma}_{ki}\rvert^{1/2}}\exp\left( -\frac{1}{2}(\B{\mu}_i-\B{\bar{\mu}}_k)^\top\B{\Sigma}^{-1}_{ki}(\B{\mu}_i-\B{\bar{\mu}}_k) \right)
\end{equation}
\noindent for $K=4$ Gaussian components, where $\pi_k$ is the fraction of stars in the $k$th Gaussian (with $\sum^K_{k=1}\pi_k=1$), $\B{\mu}_i$ the proper motion vector of the $i$th star, $\B{\bar{\mu}}_k$ the velocity centroid of the $k$th Gaussian, and $\B{\Sigma}_{ki}$ is the \rbold{covariance matrix} of the $k$th Gaussian and the $i$th star. Following \citet{clarkson2012proper} and \citetalias{hosek2015arches}, we take $\B{\Sigma}_{ki}=\textbf{S}_i + \textbf{Z}_k$ where $\textbf{S}_i$ is the (diagonal) velocity covariance matrix of the $i$th star and $\textbf{Z}_k$ the \rbold{covariance matrix} of the $k$th Gaussian fit. 

We employ Bayesian analysis to estimate kinematic parameters. For these data, the posterior probability distribution is given by
\begin{equation}
P(\B{\pi},\bar{\B{\mu}}, \textbf{Z} \rvert \B{\mu},\textbf{S}) = \frac{P(\B{\mu},\textbf{S}\rvert\B{\pi},\bar{\B{\mu}},\textbf{Z})P(\B{\pi},\bar{\B{\mu}},\textbf{Z})}{P(\B{\mu},\textbf{S})}
\end{equation}

\noindent where $P(\B{\pi},\bar{\B{\mu}}, \textbf{Z} \rvert \B{\mu},\textbf{S})$ is the posterior probability of the model, $P(\B{\mu},\textbf{S}\rvert\B{\pi},\bar{\B{\mu}},\textbf{Z})$ is the probability of the observed star velocity distribution given the model, $P(\B{\pi},\bar{\B{\mu}},\textbf{Z})$ is the prior probability of the model, and $P(\B{\mu},\textbf{S})$ is the sample evidence. 
Here, $\B{\pi}$ is the set of $\pi_k$ values, $\bar{\B{\mu}}$ the set of Gaussian velocity centroids, $\textbf{Z}$ the set of Gaussian covariance matrices, $\B{\mu}$ the set of observed stellar proper motions, and $\textbf{S}$ the set of proper motion error matrices.

\begin{figure*} \label{VPD}
\begin{center}
\includegraphics[scale=0.35]{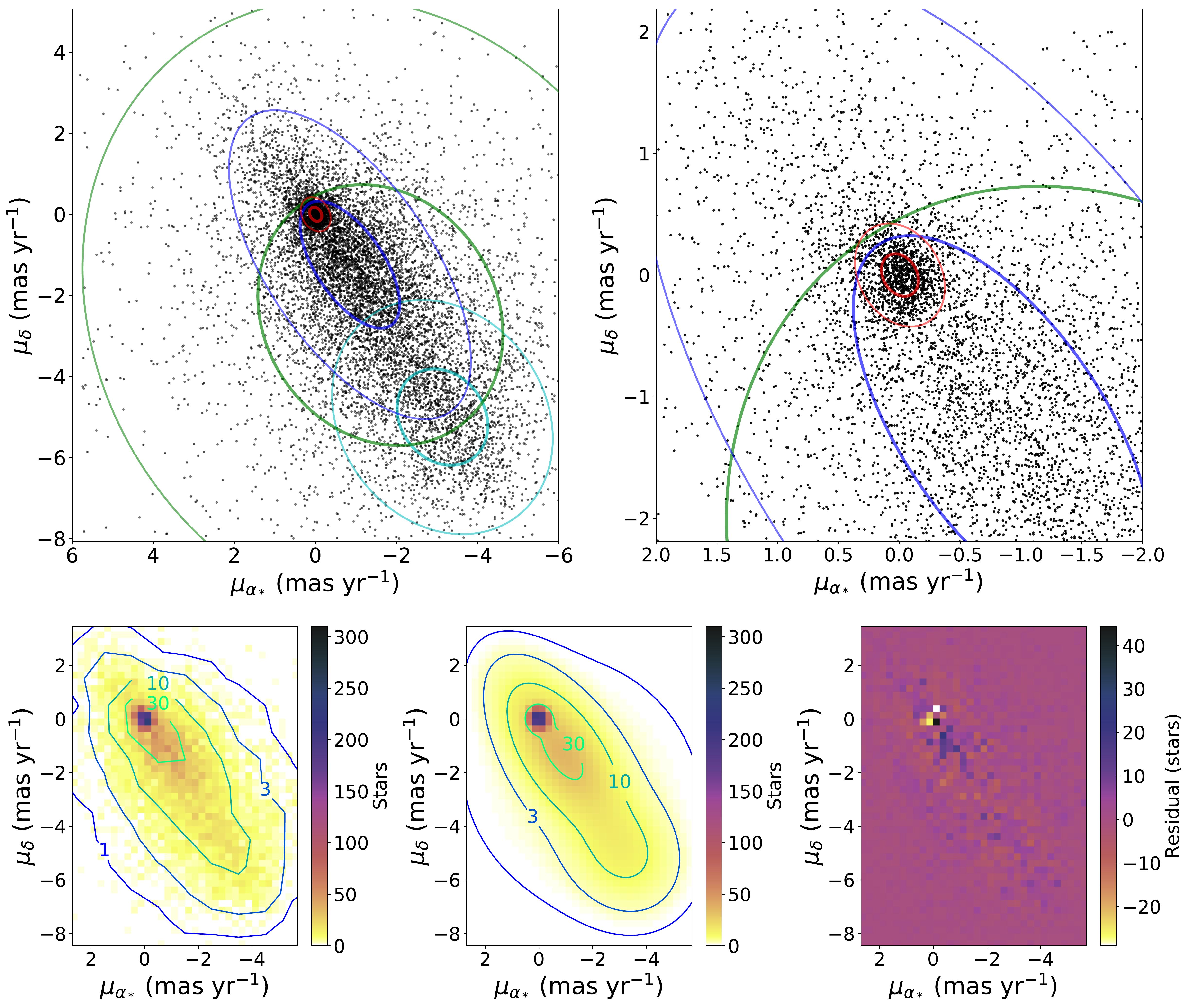}
\end{center}

\caption{\textit{Top left, top right}: Vector point diagrams of the observed proper motions and the best-fit Gaussian mixture model. The cluster Gaussian is shown in red, the first field Gaussian in blue, the second in green, and the third in cyan. The $1\sigma$ and $2\sigma$ contours are shown for each Gaussian. Note that the field Gaussians are offset from the cluster Gaussian due to differential proper motions along the line sight due to Galactic rotation. The right panel is the same as the left panel, but zoomed in on the cluster to show the cluster Gaussian. Proper motions along the galactic plane stretch from the upper left to the lower right corners of the VPDs. \rbold{\textit{Bottom}: A comparison between the observed (\textit{left}) proper motion distribution and Gaussian mixture model convolved with a Gaussian corresponding to typical proper motion errors (\textit{right}). We find that the model is in good agreement with the observed proper motions.}}
\end{figure*}

Using our Bayesian framework, we fit our Gaussian mixture model to the velocity distribution of the observed sample. Hereafter, we refer to the first Gaussian as the cluster Gaussian and the others as the field Gaussians. 
For each Gaussian, we independently vary $\pi_k$, $(\mu_{\alpha_*,k},\mu_{\delta,k})$: the mean R.A.~and Dec.~proper motions of the Gaussian, $\sigma_k$: the standard deviation of the Gaussian along its semi-major axis, $f_k$: the ratio of the Gaussian's semi-minor and semi-major axes, and $\theta_k$: the angle parameterizing the orientation of the Gaussian \rbold{clockwise} with respect to $+\alpha^*$ as shown in the top panels of Figure \ref{VPD}. 
For the position of the cluster Gaussian, we adopt a Gaussian prior. For all other parameters, we adopt uniform priors in order to be agnostic about the velocity distribution in our fit.
Our choice of $K=4$ Gaussians is justified for two reasons. First, the Gaussian mixture model accurately reproduces the empirical distribution of proper motions (bottom panels of Figure \ref{VPD}). 
Second, the Bayesian Information Criterion \citep[BIC]{schwarz1978bic} strongly prefers $K=4$ over $K=5$ ($\Delta$BIC $=21.0$) and $K=3$ ($\Delta$BIC $=61.7$).
\rbold{Interestingly, we find that the cluster's velocity distribution is inconsistent with being isotropic at the $3.6\sigma$ level, appearing elongated along the cluster's velocity vector.}

This fit to the cluster and field population kinematics was executed using the Bayesian inference tool \texttt{MultiNest}, a multi-modal nested sampling algorithm \citep{feroz2009multinest}.
We interface with \texttt{MultiNest} through the Python module \texttt{PyMultiNest} \citep{buchner2014pymultinest}.
This procedure produces a posterior probability distribution over the parameters of the Gaussian mixture model, with which we can determine a best-fit velocity distribution model.

Only stars with proper motion errors below $0.65$ mas yr$^{-1}$ and photometric errors below $0.06$ in F153M are used in the model fit. In addition, we remove two proper motion outliers which are found to significantly skew the parameters of the fourth Gaussian.
The outlier proper motions are separated from the field population by $\sim30$ mas yr$^{-1}$ and $\sim40$ mas yr$^{-1}$ and are blue foreground stars. We calculate membership probabilities for 10,543 stars, 10,541 of which are used to fit the Gaussian mixture model.
The priors and final parameters values are outlined in Table \ref{membership-table}, and the model overlaid on a scatter plot of stellar proper motions (or a vector point diagram, VPD) is shown in Figure \ref{VPD}.
The quality of our proper motion membership is readily apparent in the CMDs shown in Figure \ref{CMDs}.

\begin{table*}
%\begin{adjustwidth}{-.5in}{-.5in}
\caption{Gaussian Mixture Model of Proper Motions: Priors and Final Parameter Values}\label{membership-table}
\begin{center}
\begin{tabular}{l r r r r r r r r}
\toprule
& \multicolumn{2}{c}{Cluster Gaussian} & \multicolumn{2}{c}{Field Gaussian 1} & \multicolumn{2}{c}{Field Gaussian 2} & \multicolumn{2}{c}{Field Gaussian 3} \\
\cmidrule(lr){2-3}\cmidrule(lr){4-5}\cmidrule(lr){6-7}\cmidrule(lr){8-9} \\
Parameter & Prior & Result & Prior & Result & Prior & Result & Prior & Result \\
\midrule
\multicolumn{1}{l|}{$\pi_k$} & \U{0}{1} & $0.109\pm0.006$ & \U{0}{1} & $0.482\pm0.022$ & \U{0}{1} & $0.270\pm0.019$ & \U{0}{1} & $0.137\pm0.020$ \\
\multicolumn{1}{l|}{$\mu_{\alpha_*,k}$ (mas yr$^{-1}$)} & \G{0}{0.3} & $-0.01\pm0.01$ & \U{-6}{6} & $-0.85\pm0.06$ & \U{-6}{6} & $-1.60\pm0.10$ & \U{-6}{6} & $-3.13\pm0.13$ \\
\multicolumn{1}{l|}{$\mu_{\delta,k}$ (mas yr$^{-1}$)} & \G{0}{0.3} & $0.00\pm0.01$ & \U{-6}{6} & $-1.24\pm0.09$ & \U{-6}{6} & $-2.48\pm0.12$ & \U{-6}{6} & $-5.00\pm0.16$ \\
\multicolumn{1}{l|}{$\sigma_k$ (mas yr$^{-1}$)} & \U{0}{1} & $0.19\pm0.01$ & \U{0}{8} & $1.81\pm0.08$ & \U{0}{8} & $3.36\pm0.09$ & \U{0}{8} & $1.26\pm0.12$ \\
\multicolumn{1}{l|}{$f_k$ (mas yr$^{-1}$)} & \U{0}{1} & $0.75\pm0.07$ & \U{0}{1} & $0.46\pm0.02$ & \U{0}{1} & $0.85\pm0.03$ & \U{0}{1} & $0.82\pm0.09$ \\
\multicolumn{1}{l|}{$\theta_k$ (rad)} & \U{0}{\pi} & $1.04\pm0.16$ & \U{0}{\pi} & $0.97\pm0.02$ & \U{0}{\pi} & $0.98\pm0.10$ & \U{0}{\pi} & $0.94\pm0.26$ \\
\bottomrule
\end{tabular}
\end{center}
% Definitely want to be consistent here re: sig figs, should I just cut it off at 2 digits past decimal?

Priors and posteriors for the Gaussian mixture model fit to the cluster and field kinematics.
Free parameters of the $k$th Gaussian include: $\pi_k$: fraction of stars in the $k$th Gaussian component, $\B{\bar{\mu}}_k=(\mu_{\alpha_*,k},\mu_{\delta,k})$: proper motion centroid in R.A.~and Dec.~direction, $\sigma_k$: standard deviation along the semi-major axis, $f_k$: ratio of the semi-minor axis to semi-major axis, $\theta_k$: clockwise angle from the +$\alpha_*$~axis to the semi-major axis as shown. Here, \U{\textnormal{left bound}}{\textnormal{right bound}} denotes a uniform prior and \G{\textnormal{mean}}{\textnormal{standard deviation}} denotes a Gaussian prior.
%\end{adjustwidth}
\end{table*}

\subsection{Cluster Membership} \label{finalmembership}
\begin{figure}
\includegraphics[scale=0.33]{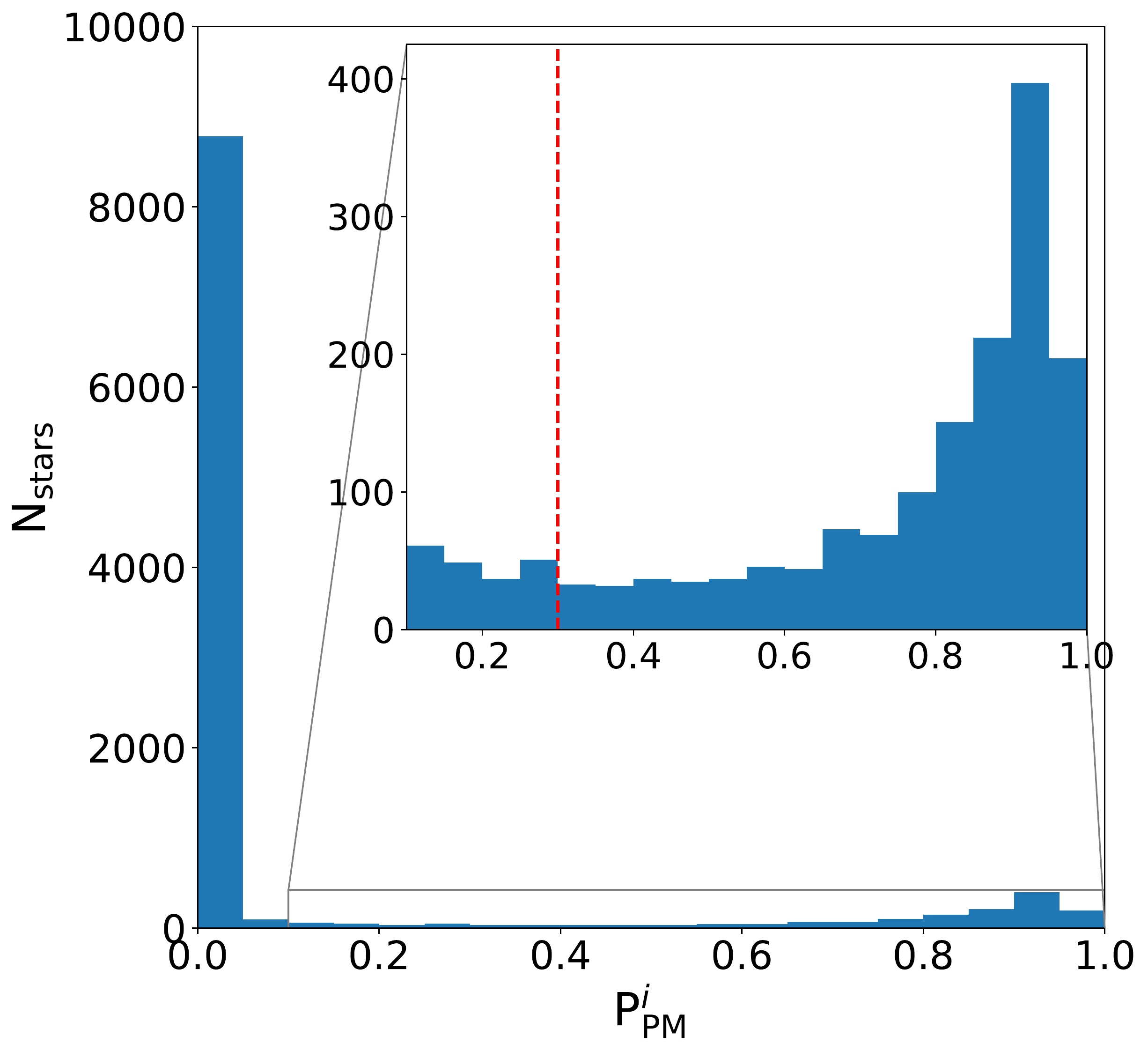}
\centering
\caption{A histogram of kinematic cluster membership probabilities $P^i_{\textnormal{PM}}$. The red dashed line indicates $\mbold{P^i_{\textnormal{mem}}}=0.3$. Stars with $P^i_{\textnormal{mem}}>0.3$ \rbold{within 3.2 pc of the cluster center} which satisfy the requisite magnitude \rbold{criterion} are included in profile fitting weighted by their membership probabilities (\S\ref{radialprofilefit}).} \label{membership-hist}
\end{figure}

Once the cluster and field kinematics are modeled using a Gaussian mixture model (\S\ref{membership}), we assess the relative probability that a given star lies within the cluster Gaussian as the \rbold{kinematic} cluster membership probability:
\begin{equation}
P^i_{\textnormal{PM}} = \frac{\pi_1P^i_1}{\sum^4_{k=1}\pi_kP^i_k}
\end{equation}

\noindent where $P^i_{\textnormal{PM}}$ is the kinematic cluster membership probability and the sum is over all Gaussians in the mixture model. The distribution of cluster membership probabilities is shown in Figure \ref{membership-hist}.

\rbold{After we calculate the kinematic cluster membership probabilities, we conservatively reject color outliers from our proper-motion selected cluster sample in order to reduce contamination from our field.} After performing a very coarse color cut, we fit a spline to high-probability kinematically determined cluster members ($P^i_{\textnormal{PM}}>0.7$) on the CMD to define a cluster sequence. An iterative procedure is used to calculate the standard deviation of the F127M-F153M and F139M-F153M colors of the cluster sequence at different magnitudes, and stars which lie $3\sigma$ from the cluster sequence are assigned a color membership probability $P^i_{\textnormal{color}}=0$. Because contamination from the RC is most prominent in the 16 to 18 mag F153M regime, we assume that the color spread of the cluster in this regime is the same as for the 15-16 mag bin. The final cluster membership probability is calculated as $P^i_{\textnormal{mem}}=P^i_{\textnormal{PM}}\times P^i_{\textnormal{color}}$ (\S\ref{membership}). Of 10,543 stars whose membership probabilities we assess, we find a total of 1,275 stars with $P^i_{\textnormal{mem}}>0.3$, 1,169 of which lie between $0.3$ and $3.2$ pc of the cluster center. By summing the membership probabilities and accounting for image and area completeness, we infer the existence of $835.5$ cluster members with F153M\textsubscript{D} magnitude brighter than $18.9$ mag lying between $0.3$ and $3.2$ pc of the center of the cluster. Sample statistics at various stages of the analysis can be found in Table \ref{catalogs}.

\begin{table*}
\caption{Sample Size with Different Criteria}\label{catalogs}
\begin{center}
\begin{tabular}{l l r}
\toprule
Catalog & Description of Cut & Stars in catalog \\
\hline
Proper motions & Required detection in 3 epochs & 51,782 \\
Proper motion error cut & $\sigma_{v_x},\sigma_{v_y}<0.65\textnormal{ mas yr}^{-1}$ & 20,802 \\
Membership catalog & $\sigma_{\textnormal{F153M}}<0.06$ mag & 10,543 \\
30\% kinematic members & $P_{\textnormal{PM}}>0.3$ & 1,463 \\
Kinematic and color membership & $P_{\textnormal{mem}}>0.3$ & 1,275 \\
Radial profile (full cluster) & F153M\textsubscript{D} $<18.9\textnormal{ mag}$, $r<3.2\textnormal{pc}$ & 715 \\
Mass-binned profiles & F153M\textsubscript{D} $<19.7$, $0.3\textnormal{ pc}<r<3.2\textnormal{pc}$ & 999 \\
\bottomrule
\end{tabular}
\end{center}

Number of stars present in the HST WFC3-IR Quintuplet data and catalogs throughout the analysis after various cuts are made. Each catalog is the number of stars satisfying the listed cut as well as those above it. The star counts shown here are not weighted by membership probability and completeness.\\
$\sigma_{v_x}$ and $\sigma_{v_y}$ refer to the proper motion errors in the $x$ and $y$ directions, respectively. \\
$\sigma_{\textnormal{F153M}}$ refers to the magnitude error in F153M. This condition is enforced separately on each epoch, and only stars that satisfy this condition in all detected epochs are retained. \\
F153M\textsubscript{D} is averaged over all epochs in which a star is detected. 
\end{table*}

\subsection{Radial Profile Fitting} \label{radialprofilefit}
To build our full radial density profile, we restrict our star sample to those which (1) have a membership probability $P^i_{\textnormal{mem}}>30\%$, (2) have F153M\textsubscript{D} $<18.9$ mag (corresponding to $\sim 30\%$ image completeness), and (3) lie within $3.2$ pc of the center of the cluster (which corresponds to $\sim 30\%$ area completeness). These cuts ensure that the profile fits are not biased by stars with large completeness corrections or small cluster membership probabilities.

Using \texttt{MultiNest}, we fit radial profiles to various models in the literature with Bayesian analysis techniques similar to those used for the kinematic membership probability analysis. We maximize the log-likelihood function $\log\mathcal{L}$ given by
\begin{equation}
\log\mathcal{L} = \sum^N_i w(r_i,m_i)\log\Sigma(r_i,\theta)
\end{equation}

\noindent
where $N$ is the number of stars used for the fit, $\Sigma(r_i,\theta)$ is a particular surface density profile (as a function of $r$ and the model parameters $\theta$, specific to the particular profile model used, and $w(r_i,m_i)$ is a weighting factor given by
\begin{equation}
w(r_i,m_i) = \frac{P_i}{A(r_i)C(r_i,m_i)}
\end{equation}

% This last sentence feels a bit vague
\noindent where $P_i$ is the membership probability of the $i$th star, $A(r_i)$ is the interpolated fraction of the total area covered by the field of view as a function of radius (taking into account regions of the detector which do not provide data, i.e. edge defects and the circular defect at the bottom of the chip), and $C(r_i,m_i)$ is the image completeness at the magnitude and radius of the $i$th star. 
See \citet{richardson2011pandas} and Footnote 3 in \citet{cicuendez2017tracing} for the derivation and further discussion of the likelihood function, as well as \citetalias{hosek2015arches} and \citet{do2013weights} for further discussion of the methodology. The radius $r_i$ is taken to be the distance to the center of the cluster, defined here as the position of the star Q12, approximately located at $\alpha(J2000)=17^{\textnormal{h}} 46^{\textnormal{m}} 15.13^{\textnormal{s}}$, $\delta(J2000)=-28\degree 49' 34.70''$ \citep{glass1990q12}.

We calculate the area completeness $A(r_i)$ by adopting the same radial bins as used in \S\ref{completeness} for the image completeness analysis. We then divide the area of the image with coverage by the theoretical area of that radial bin given an infinite field-of-view. Linear interpolation is then applied to extend the area completeness values to all radii.

In this work, we first consider the radial profile model of \citet{king1962structure}, which is widely used in the study of globular clusters \citep[e.g.,][]{trager-king}. The King model has $\Sigma(r,\theta)$ given by
\begin{equation}
\Sigma\left(r,\lbrace r_c,r_t,b\rbrace\right) =
\Sigma_0(r_c,r_t,b)\left[ \frac{1}{\left( 1 + \frac{r^2}{r_c^2} \right)^{\frac{1}{2}}} - \frac{1}{\left( 1 + \frac{r_t^2}{r_c^2} \right)^{\frac{1}{2}}} \right]^2\Theta(r_t-r) + b 
\end{equation}
\noindent parameterized by a core radius $r_c$ and tidal radius $r_t$, as well as a background term $b$ to capture remaining contaminants ($\Theta(r)$ is the Heaviside step function). The King model is characterized by a flattened core and a sharp truncation at a tidal radius $r_t$. For the King model, the model parameters are $\theta=\lbrace r_c,r_t,b\rbrace$ where $b$ is a background term accounting for field contaminants.

We also consider the Elson, Fall, and Freeman model prescribed by \citet{elson1987structure}, hereafter an EFF model, is described by $\Sigma(r,\theta)$ with
\begin{equation}
\Sigma\left(r,\lbrace\gamma,a,b\rbrace\right) = \Sigma_0(\gamma,a,b)\left( 1 + \frac{r^2}{r_c^2}\left( 2^{2/\gamma} - 1 \right) \right)^{-\gamma/2} + b
\end{equation}

\noindent so that the fitting parameters are $\theta=\lbrace\gamma,r_c,b\rbrace$. The EFF model, like the King model, has a flattened core with core radius $r_c$. However, the outer profile behavior is described as a power law with a power law slope $-\gamma$. The EFF model has been found to be a good description of a large number of YMCs and has been found to be a good fit to globular clusters in the Magellanic Clouds and other galaxies \citep{mackey_globular_a,mackey_globular_b,mackey_globular_c,McLaughlin_EFF}. In both cases, $\Sigma_0(\theta)$ is calculated analytically to yield a correctly normalized radial profile such that the integral under the radial profile is equal to the number of stars included in the fit, taking into account appropriate weighting from membership probabilities, area completeness, and image completeness.

\subsection{Mass Segregation and Cluster Asymmetry} \label{procedure-mass-seg}
In order to investigate mass segregation and the possibility of tidal tails and other asymmetric structure, we further divide the catalog into subsamples and compare their profiles as fit by the EFF model.

We partition the catalog into three bins of approximately equal total weight $\sum_iw(r_i,m_i)$.
We include stars between $0.3$ and $3.2$ pc from the cluster center with F153M\textsubscript{D} $<19.7$ mag.
Stars in the central 0.3 pc of the cluster are excluded for these mass-binned fits in order to deepen the completeness-limited F153M\textsubscript{D} depth from 18.9 mag ($\sim4.7$ M$_\odot$) to $19.7$ mag ($\sim1.9$ M$_\odot$) and allowing us to probe for mass segregation at lower masses.
This radial cut excludes 55 stars near the center of the cluster with F153M\textsubscript{D} $<19.7$ mag, leaving 999 stars.
The total weight in each mass bin, from the most massive bin to the least, is $430.0$, $452.6$, and $405.6$ stars.
The catalog is partitioned along F153M\textsubscript{D} $=17.8$ mag and $19.1$ mag.

To test for tidal tails or other structure asymmetry, we consider the same sample of stars as in our tests for mass segregation (\S\ref{masssegregation}). We draw dividing lines $\pm45\degree$ from the direction of motion of the cluster which partition the field-of-view into four regions. The two regions which overlap with the line along which the Quintuplet is moving are considered the ``parallel'' sample, and the other two are considered the ``perpendicular'' sample. To avoid an unreasonably high area completeness correction, we only include stars with a projected distance $<3.2$ pc from the cluster center. We include stars within $0.3$ pc in order to get a complete radial profile along both directions, which corresponds to a limit of F153M\textsubscript{D} $<18.9$ mag ($\mbold{\sim}4.7$ M$_\odot$). However, this reduces the total sample of stars, which limits our sensitivity in the outermost bins in particular.
Our data allow us to probe for tidal structures within the total field of view, which subtends a box with dimensions $\sim4.7$ pc $\times$ $4.7$ pc.

\section{Results}

\subsection{No Evidence for Tidal Truncation} \label{king-model}
King profiles typically provide good descriptions of the spatial distribution of stars in globular clusters, though there is some discrepancy for young clusters in the LMC \citep[e.g.,][]{mackey_globular_b}.
The King model provides a good fit to the Quintuplet core with a core radius $r_c=0.80^{+0.14}_{-0.12}$ pc; however, we find no evidence for a King-like tidal truncation out to $3$ pc (3$\sigma$ lower limit). The priors and final values for the King model profile fit can be found in Table \ref{kingfit} and full posteriors are show in Figure \ref{king_full} in Appendix \ref{summaryplots}.
\rbold{We note that the fitted value of the King core radius is not physical, since our $r_t$ posterior distribution runs up against the upper edge of the prior, forcing a larger best-fit $r_c$ (and field contamination) value to compensate.
Instead, we adopt as the fiducial core radius of the Quintuplet to be the best-fit value from the EFF profile fit as reported in \S\ref{elsonfit}.}

Based on the King model fits, we place a lower limit on the concentration parameter as $c=\log(r_t/r_c)$ at $c\gtrsim1.3$ and $c\gtrsim0.7$ for the Arches and Quintuplet, respectively.
\citet{portegies2010young} provide a relation between an EFF power law slope $\gamma$ and $c$: in the $c\rightarrow\infty$ limit, $\gamma\rightarrow2$.
A slope $\gamma\sim2$ is consistent with the Arches slope, suggesting that the Arches tidal radius, if it exists, lies very far from the center of the cluster.
The Quintuplet's steeper power law slope $\gamma\sim2.5$ perhaps indicates a very rough estimate for the true concentration parameter $\sim1.1$, putting the tidal radius at $r_t\sim 8$ pc. \rbold{Ultimately, future wide-field observations of the Quintuplet will be required to probe for a tidal truncation beyond $\sim3$ pc.}

The lack of tidal truncation is not a particularly surprising result.
While tidal truncations have been observed for many older Milky Way globular clusters, the lack of a clear tidal radius has been reported for a number of YMCs, and the non-truncated EFF model has been a sufficient description of many extragalactic star clusters \citep{mackey_globular_a,mackey_globular_b,mackey_globular_c,san2012newly,larsen2004structure,bastian2013lumprof}.
Additionally, \citetalias{hosek2015arches} ruled out the existence of an Arches tidal truncation radius out to 2.8 pc at the $3\sigma$ level, indicating that tidal truncation is not prevalent in young star clusters, even in strong tidal environments.

% Add a parameter table here
\begin{table}
\caption{King Model Profile Fit Results} \label{kingfit}
\begin{center}
\begin{tabular}{c c c c}
\toprule
Parameter & & Prior & Value \\
\hline
Core radius & $r_c$ (pc) & \U{0}{2} & 0.80$^{+0.14}_{-0.12}$ \\
Tidal radius & $r_t$ (pc) & \U{0}{4.5} & $>3^*$ \\
Field contamination & $b$ (stars pc$^{-2}$) & \U{0}{30} & 6.4$^{+1.6}_{-1.5}$ \\
Normalizing factor & $\Sigma_0$ (stars pc$^{-2}$) & --- & 283$^{+36}_{-30}$ \\
\bottomrule
\end{tabular}
\end{center}
*: $3\sigma$ lower limit for $r_t$ calculated from the posterior distribution.
\end{table}

\subsection{The Radial Profile of the Quintuplet} \label{elsonfit}
The inability to detect a tidal radius in the Quintuplet leads us to conclude that the profile near the edge of our field of view is better modeled by a power law. 
The Quintuplet's location in an extreme tidal environment could have subjected it to ongoing structural disruption, which could be manifested at the present time as a lack of a clear tidal radius.
In contrast to a King model, fitting to an EFF model (described in \S\ref{radialprofilefit}) offers two clear advantages, namely, (1) reasonable constraints on all fitting parameters and (2) better agreement with the observed outer cluster profile.

\begin{figure}
\includegraphics[scale=0.35]{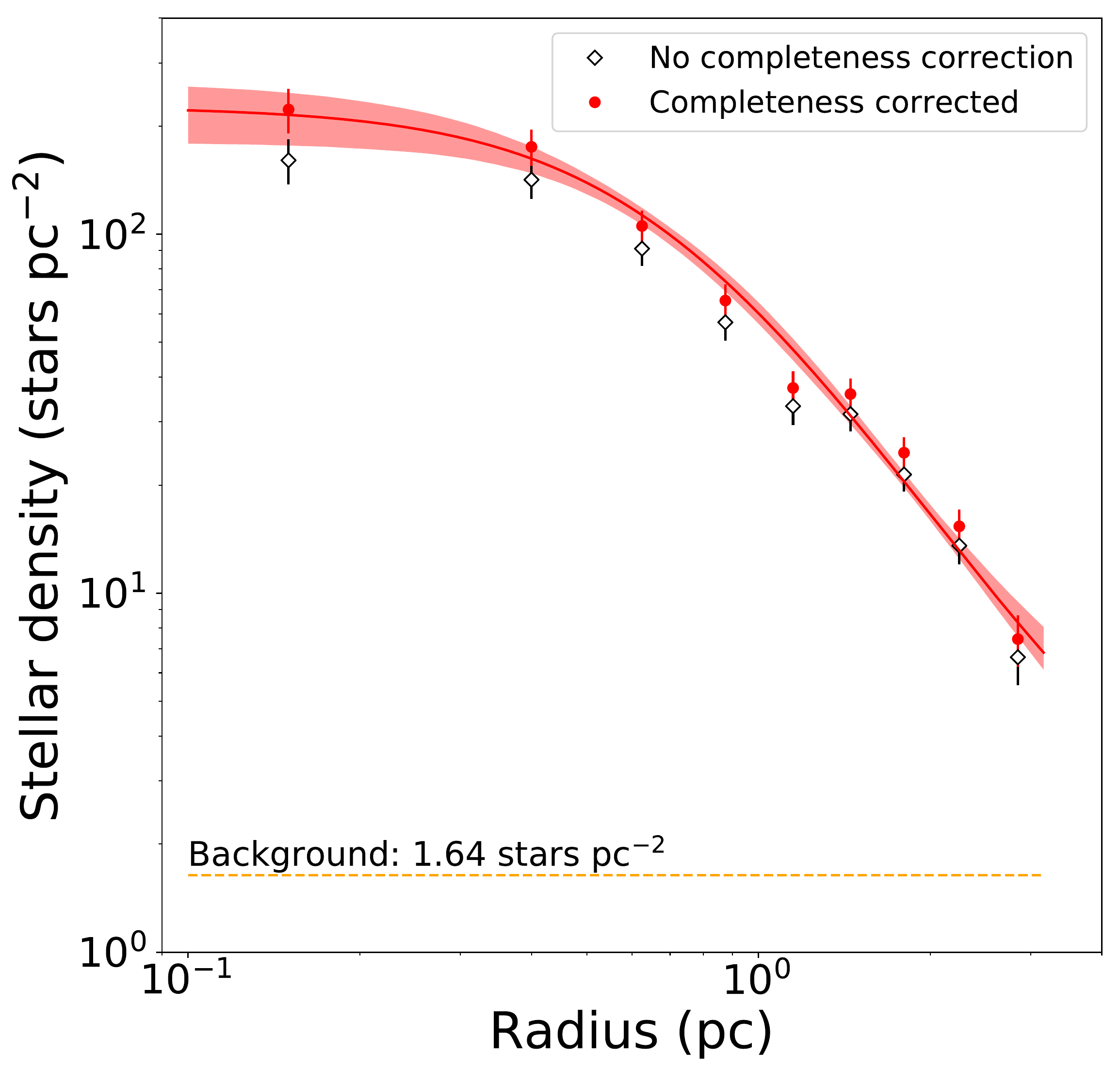}
\centering
\caption{Radial profile of the Quintuplet with EFF model fit.
The red shaded region spans the 1$\sigma$ confidence interval for values of the Quintuplet radial profile for stars with masses $\gtrsim4.7$ M$_\odot$ as determined by 1,000 radial profiles drawn from the joint posterior distribution.
The surface density at the innermost bin is shown on the data point at $r=0.15$ pc.
The open circles and red-filled circles are a binned radial profile before and after a completeness correction, respectively, though we stress that the profiles are fit using a Bayesian method which is not dependent on bin-choice.
The orange dashed line is the best-fit field contamination value for the Quintuplet, which is much smaller than the surface density.} \label{EFF-fit}
\end{figure}

The resulting best-fit cluster radius from the EFF fit is $r_c=0.62^{+0.10}_{-0.10}$ pc.
The other parameters can be found in the first row of Table \ref{EFF-table}, and the radial profile is shown in Figure \ref{EFF-fit}.
The field contamination within the central $3.2$ pc is found to be small in comparison to the size of the cluster sample itself.
For example, from the best-fit full cluster field contamination term $b=1.6$ stars pc$^{-2}$, we infer the existence of $\sim51$ field contaminant stars within $3.2$ pc with F153M\textsubscript{D} $<18.9$ mag, which constitutes $\sim6$\% of stars in the full cluster fit.

\begin{table*}
\begin{adjustwidth}{-.1in}{-.1in}
\caption{Summary of Radial Profile Fits to the EFF Model} \label{EFF-table}
\begin{center}
\begin{tabular}{l l l l c c c c c c}
\hline
& Subsample & Mass Range & Radius Range & $N_{\textnormal{stars}}^{\textnormal{a}}$ & $\gamma^{\textnormal{b}}$ & $r_c$ (pc)$^{\textnormal{c}}$ & $b$ (stars pc$^{-2}$)$^{\textnormal{d}}$ & $\Sigma_0$ (stars pc$^{-2}$)$^{\textnormal{e}}$ \\
\hline
Full cluster & & $\gtrsim4.7$ M$_\odot$ & $0 < r < 3.2\textnormal{ pc}$ & 835.5 & 2.5$^{+0.7}_{-0.4}$ & 0.62$^{+0.10}_{-0.10}$ & 1.6$^{+1.9}_{-1.2}$ & 225$^{+46}_{-37}$ \\
Split by direction & Parallel & "  " & "  " & 433.2 & 2.1$^{+0.4}_{-0.3}$ & 0.49$^{+0.11}_{-0.10}$ & 0.10$^{+0.16}_{-0.07}$ & 318$^{+90}_{-67}$ \\
& Perpendicular & "  " & "  " & 374.1 & 1.7$^{+0.6}_{-0.3}$ & 0.57$^{+0.22}_{-0.19}$ & 0.11$^{+0.20}_{-0.08}$ & 198$^{+93}_{-54}$ \\
Split by mass & High mass$^{\textnormal{f}}$ & $\gtrsim8$ M$_\odot$ & $0.3 < r < 3.2\textnormal{ pc}$ & 430.0 & $>1.5^*$ & 0.60$^{+0.18}_{-0.21}$ & 1.7$^{+1.5}_{-1.2}$ & 137$^{+107}_{-40}$ \\
& Intermediate mass$^{\textnormal{g}}$ & $\sim4.1$ to $8.0$ M$_\odot$ & "  " & 452.6 & $>1.5^*$ & 0.86$^{+0.18}_{-0.21}$ & 1.5$^{+1.5}_{-1.0}$ & 84$^{+33}_{-17}$  \\
& Low mass$^{\textnormal{h}}$ & $\sim1.9$ to $4.1$ M$_\odot$ & "  " & 405.6 & $>1.5^*$ & 1.12$^{+0.17}_{-0.19}$ & 1.6$^{+1.6}_{-1.2}$ & 50$^{+12}_{-8}$  \\
\hline
\end{tabular}
\end{center}

For each parameter, we adopt the median and the 68\% confidence interval of the 1D marginal posterior distribution.
The \textit{full cluster} and \textit{split by direction} profile fits include stars with F153M\textsubscript{D} < 18.9 mag ($M\gtrsim4.7$ M$_\odot$) within $3.2$ pc of the cluster center.
The \textit{split by mass} profile fits include stars with F153M\textsubscript{D} < 19.7 mag ($M\gtrsim1.9$ M$_\odot$) between $0.3$ and $3.2$ pc of the cluster center.
These magnitude and radial distance limits are set by image and area completeness, respectively. \\
$^{\textnormal{a}}$ Number of stars considered in the profile fit, weighted by membership probability and completeness\\
$^{\textnormal{b}}$ Outer power law slope, with a uniform prior \U{0.5}{6}\\
$^{\textnormal{c}}$ Core radius, with uniform prior \U{0}{2} \\
$^{\textnormal{d}}$ Field contamination term, with a uniform prior \U{0}{5}\\
$^{\textnormal{e}}$ Normalizing factor, solved for by setting the area integral over the EFF profile model equal to $N_{\textnormal{stars}}$\\
$^{\textnormal{f}}$ F153M\textsubscript{D} $<17.8$ mag\\
$^{\textnormal{g}}$ $17.8$ mag $<$F153M\textsubscript{D} $<19.1$ mag\\
$^{\textnormal{h}}$ $19.1$ mag $<$F153M\textsubscript{D} $<19.7$ mag\\
$^*$ $3\sigma$ lower limit\\
\end{adjustwidth}
\end{table*}

\subsection{The Search for Mass Segregation} \label{masssegregation}
\begin{figure*}
\includegraphics[scale=0.35]{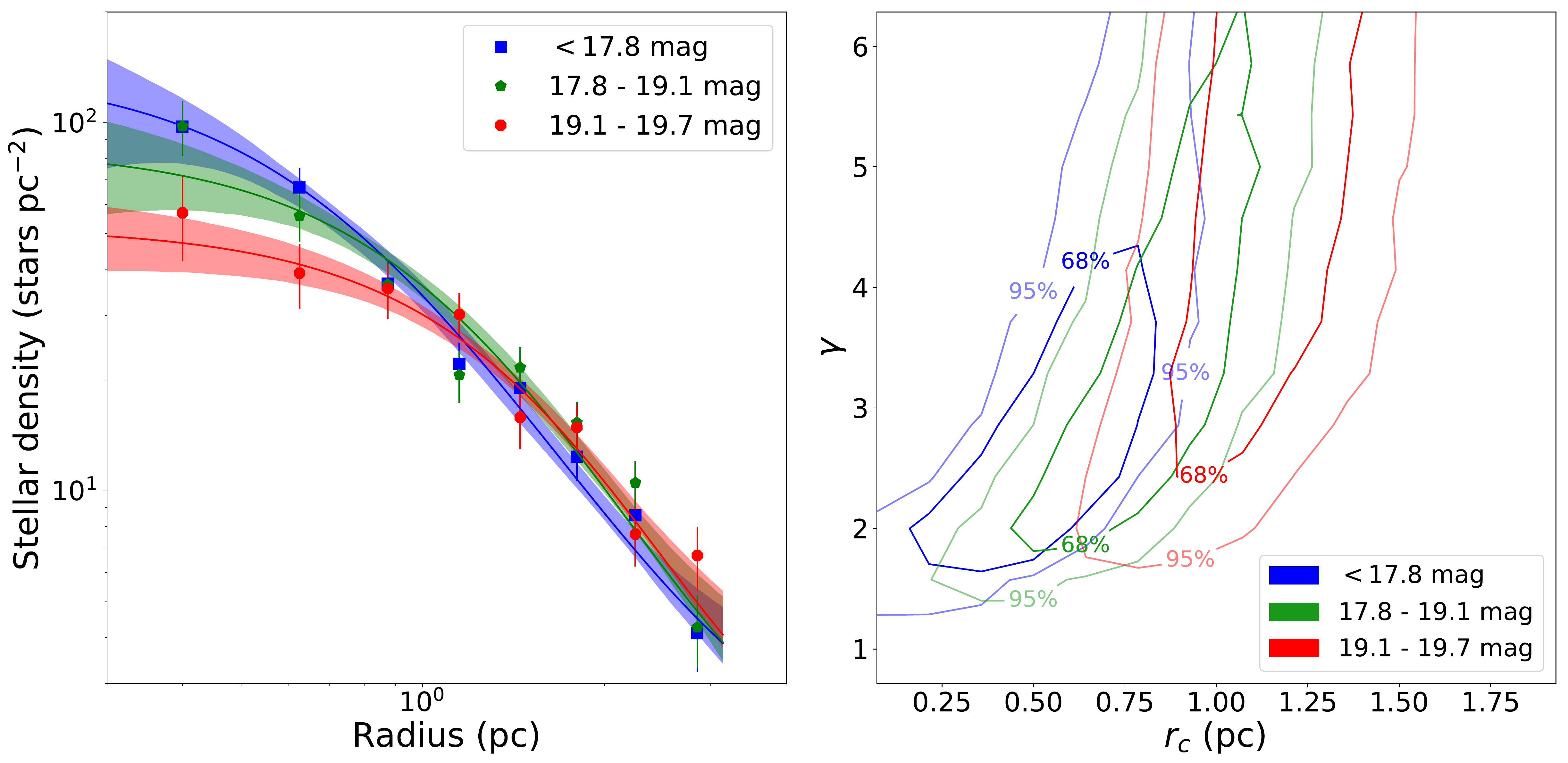}
\centering
\caption{\textit{Left}: Radial profiles for the mass-binned subsamples, as well as the best-fit EFF profiles.
\textit{Right}: The joint posterior between $\gamma$ and $r_c$ for all radial density profile model fits.
The $68\%$ and $95\%$ confidence contours are shown for each.
We find marginal evidence for mass segregation, with a Kolmogorov-Smirnov test indicating that the most massive (blue) and least massive (red) are not drawn from the same parent distribution.
We stress that the profiles are fit using a Bayesian method which is not dependent on bin-choice.} \label{mass-segregation3}
\end{figure*}

The data indicate marginal evidence for mass segregation in the Quintuplet.
\rbold{The posteriors of the most massive and least massive bins are disjoint at the 68\% confidence level but overlap at the 95\% confidence level, indicating weak evidence of mass segregation (Figure \ref{mass-segregation3}).}
In addition, a broad trend is observed in which more massive bins tend to have more compact core radii.
A Kolmogorov-Smirnov test comparing these two subsamples concludes that they are unlikely to be drawn from the same parent distribution ($3.2\sigma$).
We stress that this test is performed on the subsamples' distributions of distances from the cluster center and is completely independent of the profile fit.
Considering the overlap of the profile fit posteriors together with the Kolmogorov-Smirnov test, we conclude that there is marginal evidence for mass segregation.

While the EFF core radii of the mass-binned Quintuplet profiles differ somewhat, the outer power law slopes of all three fits are very nearly identical with little deviation (all are consistent with $\gamma\sim3$).
In contrast, \citetalias{hosek2015arches} found under a similar framework that the heaviest Arches mass bin, $M\gtrsim12$ M$_\odot$, differed in power law slope from the less massive bins by $\Delta\gamma\sim0.50-0.75$.
Clearly, the stark difference in steepness between the mass-binned profiles considered in \citetalias{hosek2015arches} for the Arches is notably absent from our data.

One na\"ively expects a larger mass segregation signal in the Quintuplet due to its more advanced age.
The larger extent of the Quintuplet as compared to the Arches means that we unfortunately have relatively little constraining power over the outer power law slope $\gamma$ for the mass-binned profiles, though we find that the lighter mass bins are consistent with higher values of $\gamma$.
This is perhaps the opposite problem in the Arches, where the very high central concentration meant that the mass-binned profiles were unconstraining on the core radius---the Arches mass-binned profiles were fit to simple power laws.
Nevertheless, \citetalias{hosek2015arches} find under this framework that the heaviest Arches mass bin, $M\gtrsim12$ M$_\odot$, differed in power law slope from the less massive bins by $\Delta\gamma\sim0.50-0.75$.
\rbold{Though we not have much constraint on $\gamma$ for the mass-binned profiles,} the stark difference in steepness between the mass-binned profiles considered in \citetalias{hosek2015arches} for the Arches is notably absent from our data.

\subsection{Lack of a Tidal Tail} \label{notidal}
\begin{figure*}
\includegraphics[scale=0.35]{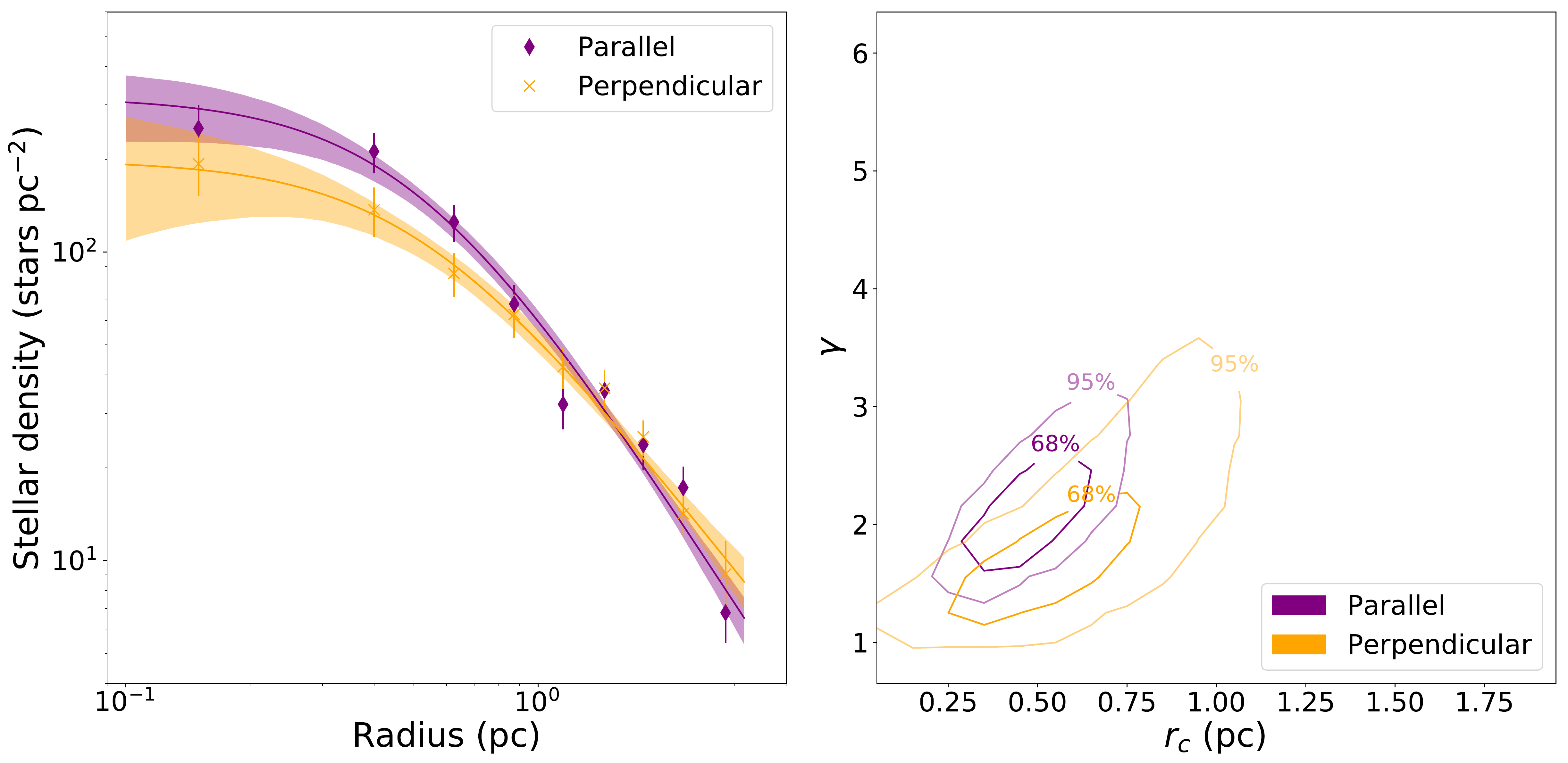}
\centering
\caption{\textit{Left}: Radial profiles for the parallel and perpendicular subsamples, as well as the best-fit EFF profiles.
\textit{Right}: The joint posterior between $\gamma$ and $r_c$ for both the parallel and perpendicular radial density profile model fits.
The $68\%$ and $95\%$ confidence contours are shown for each.
We find that the data are consistent with no cluster asymmetry or tidal tails, with the $1\sigma$ regions overlapping significantly.
A Kolmogorov-Smirnov test cannot rule out that the parallel and perpendicular profiles were drawn from the same distribution.
We stress that the profiles are fit using a Bayesian method which is not dependent on bin-choice.} \label{tidal-tails}
\end{figure*}

The posterior distributions of the parallel and perpendicular profile fits are consistent with the lack of tidal tails or any asymmetric structure.
The 68\% confidence regions for the final parameters of the two fits overlap significantly in $\gamma$--$r_c$ space, and a Kolmogorov-Smirnov test is unable to provide convincing statistical evidence for tidal tails or cluster asymmetry.
\rboldb{To check for elongation along other directions, we repeat this analysis after rotating the quadrants defining the parallel and perpendicular subsamples by angles ranging from $10\degree$ to $80\degree$ in increments of $10\degree$ and find in all cases that there is no significant cluster asymmetry.}
\rbold{$N$-body simulations of globular clusters and dwarf galaxies have shown that, for elliptical orbits, cluster tidal structures outside of a few tidal radii are expected to align along the orbit at perigalacticon and towards the GC at apogalacticon \citep{montuori2007tidal,klimentowski2009orientation}.
At distances closer to the cluster, the tidal tail is instead expected to point in the direction of the GC \citep{montuori2007tidal}.
Since the Quintuplet orbits roughly on-disk, the projection of the tidal tail on the sky should be in the same direction in either case.}
Curiously, however, the little asymmetry that we observe trends in the opposite direction, with the cluster appearing slightly wider perpendicular to the direction of motion.
The fitted profiles, as well as the fit posteriors, are shown in Figure \ref{tidal-tails}.

\subsection{A Second Red Clump?} \label{2rc}
On the CMD of the observed field (Figure \ref{unsharp_mask}, left panel), we observe an overdensity of stars approximately half an F153M magnitude dimmer than the RC which is similarly smeared along the reddening vector (Figure \ref{unsharp_mask}).
In order to roughly estimate the prominence of this overdensity, we calculate a number ratio between the number of stars in the overdensity and the number of stars in the RC.
We assume here conservatively that the RC has an F153M magnitude half-width of 0.3 mag (a criterion which is also used in the extinction analysis), that the overdensity has an F153M magnitude half-width of 0.15 mag, and that the centers of the F153M magnitudes of the two populations are separated by 0.5 mag.
We then find that the number ratio between the overdensity and the RC is $n_{\textnormal{feature}}/n_{\textnormal{RC}}\simeq0.36\pm0.02$.
There does not appear to be a significant difference in the spatial or velocity distributions of the two RC populations.

We first consider the hypothesis that the mysterious overdensity is the ``faint RC'' (fRC) reported by \citet{MZdoubleRC}, which is generally attributed to an X-shaped structure in the bulge \citep[e.g.,][]{ness2016x}.
It has widely been observed that the magnitude difference between the ``bright RC'' (bRC) and fRC decreases as the galactic latitude approaches the plane, with the $K$-band magnitudes of the two features converging at $|b|\simeq4\degree$ \citep{saitoxshapedbulge}.
Physically, it is argued that, as one approaches the midplane, the difference in distance between the arms of the X-shape decreases, so that the double RC should merge into a single RC at low latitudes.
However, in this picture, the Quintuplet, which lies almost perfectly in the midplane, should not express a double RC---we nevertheless observe that such a feature exists ($|b|\approx0.06\degree$).
We conclude, therefore, that the secondary RC population observed in our data is inconsistent with being due to an X-shaped bulge which merges spatially at low latitudes.
More photometric observations of the RC population at low latitudes would certainly elucidate whether the morphology of the bulge could really give rise to the observed feature.

More recently, it has been hypothesized that the double RC arises from two distinct RC populations, where the fRC is a He-normal, first-generation population and the bRC is a He-enhanced, later generation population \citep{lee2016rc,joo2017rc,lee2018rc}, though not without contention.
\citet{gonzalez2015reinforcing} show that this scenario requires a contrived spatial distribution to explain the double RC's latitude-dependence, calling into question the plausibility of this explanation.
Spectroscopic analysis of this secondary population would be helpful in determining whether the two RC features are distinct in composition.

We consider a third possibility, that this feature is actually the bulge's red giant branch bump population (RGBB), which \citet{natafRGBB} claim to observe at lower latitudes.
Stars ascending the red giant branch reach a stage where their hydrogen-burning layer comes in contact with the convective layer, allowing the star to dim momentarily.
This causes ascending stars to pass the same magnitude range three times, leading to an overdensity at the RGBB.
\citet{natafRGBB2013} report that the number ratio between the bulge RGBB and RC populations of $n_{\textnormal{RGBB}}/n_{\textnormal{RC}}=0.201\pm0.012$ which is around half of our observed value.
Notably, though, by examining the stellar populations of globular clusters, they find a strong correlation between metallicity and $n_{\textnormal{RGBB}}/n_{\textnormal{HB}}$, the RGBB to horizontal branch number ratio, where the horizontal branch is a population II analogue to the RC in the bulge.
Our measured value of $n_{\textnormal{RGBB}}/n_{\textnormal{RC}}$ appears somewhat consistent with values of $n_{\textnormal{RGBB}}/n_{\textnormal{HB}}$ at the high end of their observed scatter for [M/H] $\gtrsim-0.5$---with respect to globulars in their sample with [M/H] $>-0.5$, our number ratio lies above their average number ratio by one standard deviation of the globulars' $n_{\textnormal{RGBB}}/n_{\textnormal{HB}}$ values.
It remains to be seen whether the secondary population that we observe is truly consistent with the RGBB of the bulge, given the discrepancy in observed number ratio.

One final hypothesis is that the second RC is due to a bimodality in extinction due to two RC populations on either side of the GC.
Since the two populations are separated in magnitude by $\sim0.5$ mag, if one assumes a distance of $8$ kpc to the primary RC, the secondary RC would be at a distance of $\sim10$ kpc.
\citet{gonzales2018trace} use VVV data to argue that the claimed RGBB observed by \citet{natafRGBB} is actually due to RC stars tracing a spiral arm on the far side of the bulge.
Note, however, that \citet{gonzales2018trace} invokes the presence of an RGBB population to obtain better agreement with the data at $l\approx0\degree$, where a background RC population alone is insufficient to match observations; this important caveat is specifically relevant to Quintuplet \rbold{which lies at this longitude}.
While a promising conjecture, future work with Gaia DR2 will illuminate whether this proposed structural component is consistent with our data.

\section{Discussion}

\subsection{Evolution and Lifetime of YMCs in the GC}
\begin{figure}
\includegraphics[scale=0.3]{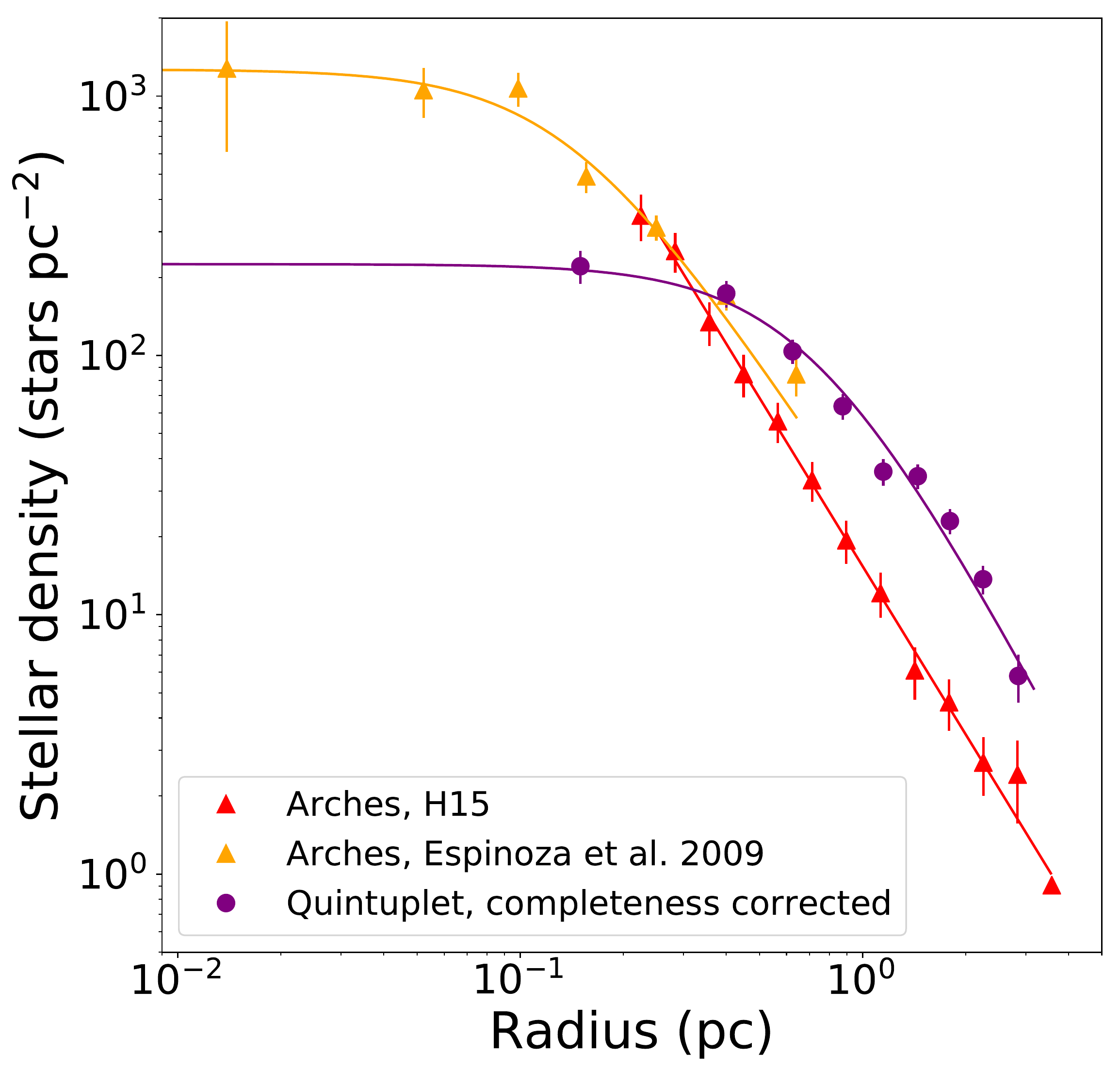}
\caption{The Arches and Quintuplet radial profiles. \citet{espinoza2009massive} provide the inner radial profile, fitting the cluster to the surface density profile prescription for the King model core: $\Sigma(r,r_c)=\frac{\Sigma_0}{1+(r/r_c)^2}$.
The outer radial profile of the Arches down to $\mbold{\sim}4.7$ M$_\odot$ is fit to a power law, where the profile is fit to a power law model: $\Sigma(r,\Gamma,b) = \Sigma_0r^{-\Gamma} + b$.
The inner Arches profile is rescaled to be continuous with the \citetalias{hosek2015arches} Arches outer profile at $0.25$ pc.
The Quintuplet profile and Arches \citetalias{hosek2015arches} profile are background subtracted.
The smaller core and higher central surface density of the Arches is apparent.} \label{compareprofiles}
\centering
\end{figure}

The Quintuplet's dynamical structure is particularly instructive in comparison to that of the Arches cluster, which, at a somewhat younger age of $\sim2.5$--$3.5$ Myr, is thought to be a ``younger cousin'' of the Quintuplet \citep{figer2002arches,schneider2013ages}.
The Arches has a similar mass and is similarly situated, lying about $\sim26$ pc in projection from the GC. 
Simulations of the orbits of the Arches and Quintuplet suggest that the two clusters may have originated in the same region of the central molecular zone (CMZ) if the Quintuplet lies in front of the GC today \citep{stolte2014orbital}.
In addition, gas cloud orbital simulations suggest that both the Arches and Quintuplet could have formed in sequence in a gas ring or stream(s) at the outer edge of the CMZ \citep{kruijssen2014clouds}.
We can view the Quintuplet and Arches as being two snapshots in the evolution of a dense cluster in a strong tidal field, albeit with the caveat that we do not know whether the clusters formed with similar initial profiles.
The difference in cluster extent is readily apparent in the Arches and Quintuplet profiles (Figure \ref{compareprofiles}).

In this context, the Quintuplet's comparatively large core radius of 0.62 pc compared to that of the Arches \citep[$0.13\pm0.03$ pc,][]{espinoza2009massive} provides some of the most direct empirical evidence to date that YMCs in the GC dissolve on a timescale of $\sim10$ Myr, as predicted by several \rbold{Fokker-Planck and} $N$-body simulations \rbold{\citep{kim1999evaporation,kim2000dissolve,pz2002dissolve}}.
The half-light radius of the Quintuplet is $1.05$ pc, which is also much larger than that of the Arches (0.48 pc, \citetalias{hosek2015arches}).
The Quintuplet is still remarkably compact for its location, with a core radius comparable to those of other YMCs which are not under the influence of such a strong tidal field, such as Westerlund 1 with a core radius of $\sim0.40$ pc.
The discrepancy in tidal forces between the two clusters can be quantified by their theoretical tidal radii.
Assuming that the Quintuplet is between 50 and 150 pc from the GC, the cluster encloses a galactic mass between $4.9\times10^6$ and $1.1\times10^8$ M$_\odot$ \citep{launhardt2002galaxy}.
With a cluster mass $\sim10^4$ M$_\odot$, this yields a theoretical tidal radius between $\sim4.4$ \rbold{and} $\sim4.7$ pc.
In contrast, Westerlund 1, at $\sim2.6$ kpc from the GC \citep{kothes2007distance}, encloses $\sim2.3\times10^{10}$ M$_\odot$.
With a cluster mass of $(4.4$--$5.7)\times10^4$ M$_\odot$ \citep{andersen2017wd1}, the theoretical tidal radius of Westerlund 1 is significantly farther out, between $\sim22$ and $\sim24$ pc.
The Quintuplet core radius is also smaller than the vast majority of extragalactic globular clusters in the seminal Mackey and Gilmore catalogs \citep{mackey_globular_a,mackey_globular_b,mackey_globular_c}.

Further observations at large radii will be required to directly measure the Quintuplet's tidal radius and elucidate the effect of a strong tidal field on a star cluster's spatial extent and gravitational influence.
If the Quintuplet tidal radius truly lies $\gtrsim4.4$ pc from the cluster center, it is unsurprising that we do not observe a tidal truncation within $\sim3$ pc.
Future wider-field observations of the Quintuplet will be necessary to corroborate or else empirically revise this estimate.

The core radii of star clusters are thought to expand as the clusters age, though the expansion rate is dependent on the initial mass function slope \citep{EFL1989slope,mackey_globular_a,mackey_globular_b,mackey_globular_c}. The initial mass function slope of the Arches has been shown to be top-heavy \rbold{\citep{hosek2019imf}}, and the same has been hinted at in the Quintuplet \citep{hussmann2012pdmf}. A difference in the initial mass function between the two clusters might explain the difference in the core radii.

\citet{park2018starclusters} conduct $N$-body simulations of star clusters in strong tidal fields with various initial conditions, integrating for 2 Myr.
They find that, in order for an Arches-like cluster with spherical symmetry to survive for $\sim2$ Myr, it must have had an initial size of $\lesssim2$ pc at a distance of 30 pc to the Galactic Center or $\lesssim23$ pc at $100$ pc.
Notably, many plausible formation scenarios they explore are of star clusters which are ``well-contained,'' that is, their initial characteristic radius is comfortably below their initial tidal radius.
As the Quintuplet has survived for $\sim5$ Myr, these constraints are also applicable.
We infer that the Quintuplet was born with a compact core radius of $\lesssim2-3$ pc in order for it not to have been torn apart by tidal forces by the present day.

\subsection{Mass Segregation in YMCs}
By examining the radial profiles of mass-partitioned subsamples of the Quintuplet catalog, we found marginal evidence of mass segregation in the Quintuplet.
This is somewhat consistent with the result of a previous study claiming more definitive evidence of mass segregation in the Quintuplet using pixel intensity histograms \citep{shin2016quintuplet}.

Nevertheless, it appears clear that mass segregation in the Quintuplet, if it exists, is less substantial than it is in the Arches \citep{hosek2015arches}.
This is quite an intriguing result---dynamical mass segregation is known to segregate clusters over a relaxation time, and the Quintuplet, as the older of the two, would na\"ively be expected to be more mass segregated than the Arches.
Though it is true that the larger size of the Quintuplet makes the outer profile more difficult to characterize than that of the Arches, we find that even a comparison of the core radii of different stellar mass bins is inconclusive.
We thus postulate several physical reasons why the Quintuplet might be less mass segregated than the Arches.

One possibility is that the highest mass stars in the high-mass bin have already gone supernova, which would naturally lead to a deficit of the most centrally-concentrated stars in the high-mass bin ($>8$ M$_{\odot}$).
Assuming a cluster age of 5 Myr, stars with initial masses larger than 44 M$_{\odot}$ are expected to have already been removed by this mechanism. 
If we adopt a top-heavy initial mass function similar to what has been observed for the Arches cluster \rbold{\citep[$dN/dm\propto m^{-1.8}$,][]{hosek2019imf}}, then this would only correspond to 16\% of the stars in this bin. 
This is insufficient to evolve an Arches-like profile to the observed Quintuplet profile, which would require a depletion of at least a factor of 2. 

%First, since the Quintuplet is somewhat older than the Arches, high mass stars would have already  one supernova at the present day.
%This would, in theory, lead to a deficit at the high mass end of the cluster's mass function.
%However, a quick back-of-the-envelope calculation suggests that only stars with masses $\gtrsim21$ M$_\odot$ would have had time to go supernova.
%\textcolor{red}{nicholas: this is like a fourth of the stars in the top bin!}

It is possible that the Arches formed with a higher degree of initial mass segregation than does the Quintuplet.
This would not be unprecedented; \citet{dib2018star} find that the extent to which mass segregation is present in dense molecular cloud cores within young star-forming regions (Taurus, Aquila, Corona Australis, and W43) positively correlates with star formation rate. 
If the Arches formed faster than the Quintuplet, then it may be expected that the Quintuplet would have less mass segregation. However, it is unclear why two such clusters in close proximity would have such different star formation rates. 

%\citetalias{hosek2015arches} find that the Arches is likely mass segregated down to around $\sim13$ M$_\odot$.
%However, they are unable to confirm mass segregation below this mass limit, leaving open the possibility that the Arches has not mass segregated much below $13$ M$_\odot$.
%More definitive evidence that low mass stars in the Arches are not mass segregated would suggest that mass segregation in the Arches is mostly dynamical rather than primordial.
%A related possibility is that the Arches cluster has a smaller star formation efficiency per free fall time, which would theoretically increase the mass segregation that the Arches would be able to attain by the present day, though this effect would have to overwhelm the presumably greater dynamical mass segregation in the Quintuplet due to its more advanced age \citep{farias2018star}. \textcolor{red}{[JLU: It isn't obvious why there is a physical connection between star formation efficiency per free fall time and mass segregation. Can we discuss? ]}

In addition, it is possible that there is some discrepancy between the dynamical and relaxation times of the two clusters, i.e., that the Arches is dynamically older than the Quintuplet.
The two clusters have a comparable number of stars, thus the relative degree of mass segregation depends only on the relaxation timescale and one expects the dynamically older cluster to be more mass segregated. \rboldb{For the Quintuplet, we assume a cluster mass of $2\times10^4$ M$_{\odot}$, a half-mass radius\footnote{We assume that the half-light radius is equal to the half-mass radius, and that the true half-light radius is 4/3 times larger than the observed (i.e. projected) half-light radius \citep[e.g.][]{Spitzer:1987}.} of 1.4 pc, and an IMF consistent with local star forming regions for 0.08 M$_{\odot}$ $<$ M $<$ 0.5 M$_{\odot}$ \citep[$dN/dm\propto m^{-1.3}$; ][]{Kroupa:2002lq}, and the Arches cluster for M $>0.5$ M$_{\odot}$ \citep[$dN/dm\propto m^{-1.8}$; ][]{hosek2019imf}. This yields an overall half-mass relaxation time of $\sim42$ Myr and a relaxation time of $\sim6$ Myr for stars around 10 M$_{\odot}$ \citep{Spitzer:1987,Spera:2016,Angelo:2018}.
The corresponding calculation for the Arches yields an overall half-mass relaxation time of $\sim13$ Myr and a relaxation time of only $\sim2$ Myr for 10 M$_{\odot}$ stars. }
Although these simplistic calculations depend on the assumption that the cluster is in virial equilibrium, it appears that the Arches is dynamically older---and thus plausibly more mass segregated---than the Quintuplet.
However, since the Arches is also more compact, this explanation is only credible if one believes that the Arches was formed as a more compact cluster than the Quintuplet.
To fully interpret the past dynamical evolution of these two clusters, we need the velocity dispersion profiles of both the Quintuplet and Arches \cite[see][for core dispersion]{clarkson2012proper}.

Tidal perturbations may also impact the mass segregation signature of a cluster. 
For example, $N$-body simulations by \citet{webb2018} suggest that tidal fields in cuspy galaxies minimize the ability of a star cluster to expand, thus causing it to have a longer dynamical and relaxation time and, thus, a lower degree of mass segregation.
Also, tidal stripping preferentially removes lighter stars from the outskirts of the cluster, \rbold{possibly reducing the observational signature of mass segregation in tidally perturbed clusters.}
$N$-body simulations have suggested that such tidal stripping effects could be an important factor in Quintuplet-like clusters which are born near the GC or have recently undergone periapse passage \rbold{\citep{baumgardt2003,portegies2004life}}.
In this scenario, one expects to observe some evidence of tidal stripping in the form of tidal tails or other cluster asymmetry.
The lack of detection of a tidal tail complicates the mass segregation story, as it is more difficult to point to tidal stripping as a mechanism for weaker mass segregation in the Quintuplet compared to the Arches.
$N$-body simulations by \citet{park2018starclusters} also show very clear tidal tails, although the authors find that, after restricting to stars above $2.5$ M$_\odot$, these tidal tails are barely visible.
It is thus not surprising that we are unable to detect a tidal tail in either the Arches \citepalias{hosek2015arches} or Quintuplet by probing with stars above $\sim2.5$ M$_\odot$ and $\sim4.7$ M$_\odot$, respectively.
The future of this hunt for a tidal tail in the Quintuplet thus lies either in probing more deeply or in looking for larger scale tidal structures.
The next generation of wide-field, high-resolution infrared instruments, such as JWST and WFIRST, are promising candidates to extend the search for a Quintuplet tidal tail.

It may be useful to look towards YMCs far from the GC to build a picture of star cluster mass segregation. Though the Quintuplet and Arches have seemingly direct age and mass analogues outside the GC in Westerlund 1 and Westerlund 2, respectively, the picture is muddied by cluster-specific details and histories. For example, \citet{andersen2017wd1} find evidence for mass segregation in Westerlund 1 for stars with masses $\gtrsim2$ M$_\odot$.
They do not detect mass segregation for stars lighter than this, although this could be a central-cluster completeness issue rather than hard evidence for its absence.
In any case, the presence of \rbold{strong} mass segregation in Westerlund 1 as compared to \rbold{weaker mass segregation in the Quintuplet} could suggest that tidal stripping is a plausible explanation, though this must be heavily \rbold{caveated} with other confounding influences as described above.
\rbold{Interestingly, by relevant timescales, one would expect the Quintuplet to be more mass segregated than Westerlund 1, whose half-relaxation time \citet{brandner2008wd1} estimate as $\sim400$ Myr, so it is quite plausible that our lack of detection is due to a suppressed mass segregation signature rather than an absence of mass segregation itself.
However, as \citet{brandner2008wd1} probe for mass segregation through the mass function rather than the profile, it is unclear whether this discrepancy could be due to differing methodologies.}
The situation in Westerlund 2 may illuminate the discussion; \citet{zeidler2017wd2} find mass segregation in Westerlund 2, but argue that the timescale for dynamical mass segregation implies that it must be primordial, though it has been proposed that mass segregation can even develop over a timescale $\lesssim1$ Myr \citep{allison2009ms,dominguez2017ms}.
NGC 3603, another YMC with an age similar to that of the Arches, also has been affected by mass segregation, although it is less clear whether or not its mass segregation is primordial in origin \citep{harayama2008ngc3603}.
Clearly, more dynamical information needs to be gathered before more definitive statements can be made about the cause of the weaker mass segregation in the Quintuplet.

\subsection{2D Cluster Bulk Motion}
Previous studies have explored possible formation sites of the Quintuplet cluster.
In particular, \citet{stolte2014orbital} run orbital simulations of the Quintuplet at various line-of-sight distances using a 2D projected velocity of $132\pm15$ km s$^{-1}$\rbold{, derived by comparing the cluster's proper motion with that of their observed field population}.
\rbold{Using a catalog containing significantly many more field stars over a larger field of view fit to a more sophisticated kinematic model} (\S\ref{membership}), we derive the 2D projected cluster bulk motion with respect to the field by taking the average difference between the center of the cluster Gaussian and the center of each of the field Gaussians weighted by the error in the center of each field Gaussian and the fraction of field stars in that Gaussian, i.e. $\frac{\pi_k}{\sum^4_{n>1}\pi_n}$. We find the 2D projected velocity to be $2.62\pm0.19$ mas yr$^{-1}$ or $102.8\pm7.6$ km s$^{-1}$ assuming a distance of 8 kpc to the Quintuplet.
This cluster proper motion \rbold{relative to the field} is slightly different than the value used by \citet{stolte2014orbital}'s simulation. \rbold{In a matched sample across the two catalogs, we find that the kinematic reference frames are the same within 0.2 mas yr$^{-1}$ or 7.8 km s$^{-1}$; thus, the difference is likely attributed to our larger sample of field stars rather than differences in the reference frame.}
\rbold{We leave it to a future work to establish the Quintuplet's absolute proper motion and orbit using a combination of Gaia and HST.}

\section{Conclusion}
Using multi-filter observations from the \textit{Hubble Space Telescope} WFC3-IR camera over a six year baseline, we produce a proper-motion catalog of Quintuplet and field stars.
We use 10,541 stars with proper motion errors $<0.65$ mas yr$^{-1}$ to model the kinematic structure of the cluster and field populations using a 4-Gaussian mixture model fit using Bayesian inference techniques.
This model is used to derive proper motion-based cluster membership probabilities for 10,543 stars, 1,275 of which are determined via kinematics and photometry to be likely bound to the Quintuplet.
This study covers a field of view 19 times larger than similar studies in the past and represents a significant improvement over previous proper motion surveys.
In order to mitigate the effect of the extreme degree of differential reddening associated with the galactic bulge field, we derive an extinction map using RC stars by tracing back their reddening vector and deriving their $A_{\textnormal{Ks}}$ extinction values.
We find that the Quintuplet cluster is reddened to $A_{\textnormal{Ks}}=2.12$ \rbold{mag}.
We combine this with an extensive image completeness analysis involving the planting and recovery of artificial stars across the field.

Taking advantage of our large field of view, we calculate the radial density profile for the Quintuplet cluster out to $3.2$ pc for stars with masses $\gtrsim4.7$ M$_\odot$ and cluster membership probabilities $P^i_{\textnormal{mem}}>0.3$, weighting by cluster membership probabilities and completeness.
Fitting the profile to King and EFF models, we find that (1) the King core radius of the Quintuplet is $0.62^{+0.10}_{−0.10}$ pc, and that (2) it is very unlikely that there is a King-like tidal truncation in the Quintuplet within $3$ pc, with the EFF model better parameterizing the outer power law slope of the Quintuplet profile.
By examining the radial profiles of mass-binned subsamples, we find marginal evidence for mass segregation.
By comparing the outer power law slopes of Arches and Quintuplet mass-binned subsamples, we conclude that mass segregation in the Quintuplet is weaker than in the Arches.
This dramatic difference could arise from a number of physical causes.
One intriguing possibility is from tidal stripping, \rbold{which could potentially reduce a mass segregation signature in a strongly tidally perturbed cluster.}
However, we are unable to detect a tidal tail, although it is quite possible that it is either very diluted, only prominent outside our field of view, or both.

Nevertheless, the Quintuplet is obviously much larger in extent than the Arches, though comparable in size with other YMCs.
All else being equal, the Quintuplet appears to be an enlarged version of the Arches and a fine specimen for studying the dynamical evolution of YMCs in a strong tidal field on a few Myr timescale.
Assuming that the Arches and Quintuplet formed with similar core radii, the lack of a tidal truncation within $\sim3$ pc for both clusters as well as the inflated core radius of the Quintuplet compared to that of the Arches provides very direct evidence for rapid YMC dynamical structure evolution near the GC, in accordance with $N$-body simulations \citep{pz2002dissolve,park2018starclusters}.
\rbold{An intriguing alternative is that these structural differences may be primordial, pointing towards different initial conditions in the two clusters.}
However, wider-field studies of the Arches and Quintuplet will be needed to determine whether there is a tidal tail signal at larger distances from the cluster center.
The calculation of the velocity dispersion and dispersion profile of the Quintuplet would provide further valuable dynamical information to elucidate this Myr evolution.
This calculation is left to a future paper.

Finally, together with recent advances in models of the Milky Way \citep[e.g.,][]{portail2017model} and unprecedented understanding of the Milky Way population from Gaia, our refined value for the proper motion of the Quintuplet warrants an updated Quintuplet orbital simulation in order to more accurately localize the birthplace of the Arches and Quintuplet.
While there are competing theories for how the Arches and Quintuplet formed, none have been confirmed with much certainty.
A determination of the birthplace and birth mechanism of the Quintuplet would provide valuable insight regarding the birth of star clusters, especially in extreme environments where the formation mechanism is far from certain.
A convincing answer to the question of how these YMCs formed will undoubtedly accompany further insight into stellar formation near the GC.

\acknowledgments
M.R.M. and A.M.G. is supported by the NSF grants AST-0909218 and AST-1412615.
A.M.G. is also supported by the Lauren Leichtman \& Arthur Levine Chair in Astrophysics.
This work is based on observations made with the NASA/ESA Hubble Space Telescope, obtained the Mikulski Archive for Space Telescopes at the Space Telescope Science Institute, which is operated by the Association of Universities for Research in Astronomy, Inc., under NASA contract NAS 5-26555.
These observations are associated with programs 11671, 12667, 12318, and 14613. \rbold{We thank the anonymous referee for their thorough reading and insightful comments which improved the content and presentation of this paper.}
% Other things
%
%
%
%
%
% %%%%%%%%%%%%

%\facility{facility ID}
\facility{HST (WFC3-IR)}
\software{Astropy \citep{astropy_collab}, IPython \citep{ipython}, Matplotlib \citep{hunter2007matplotlib}, NumPy \citep{numpy}, SciPy \citep{jones-scipy}, MultiNest \citep{feroz2009multinest}, PyMultiNest \citep{buchner2014pymultinest}}

\bibliographystyle{yahapj}
\bibliography{main}

\appendix
\section{Summary Plots for the Quintuplet catalog} \label{summaryplots}
\begin{figure*}[h!] \label{compareerrors}
\includegraphics[scale=0.3]{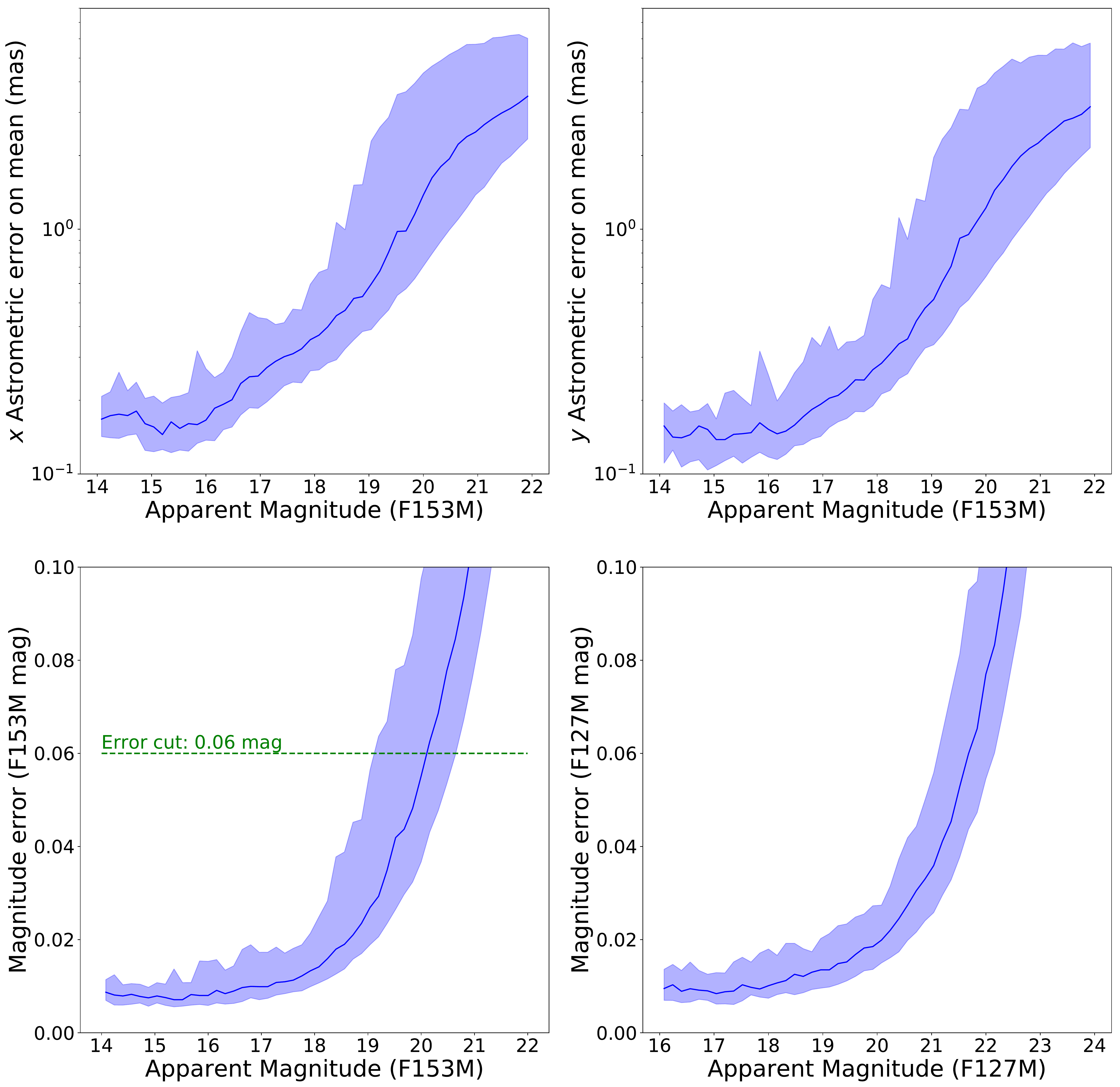}
\centering
\caption{\textit{Top}: Median astrometric errors versus F153M magnitudes on the mean.\\
\textit{Bottom}: Median RMS magnitude errors versus F153M magnitudes for Quintuplet stars that we have proper motions for, with the 16th through 84th percentiles represented in the shaded region. The green dashed line shows the F153M magnitude cut of 0.06 mag applied in this analysis for the Gaussian mixture model of the cluster/field kinematics (\S\ref{membership}).\\}
\end{figure*}

\begin{figure*}[h!] \label{chisq}
\includegraphics[scale=0.3]{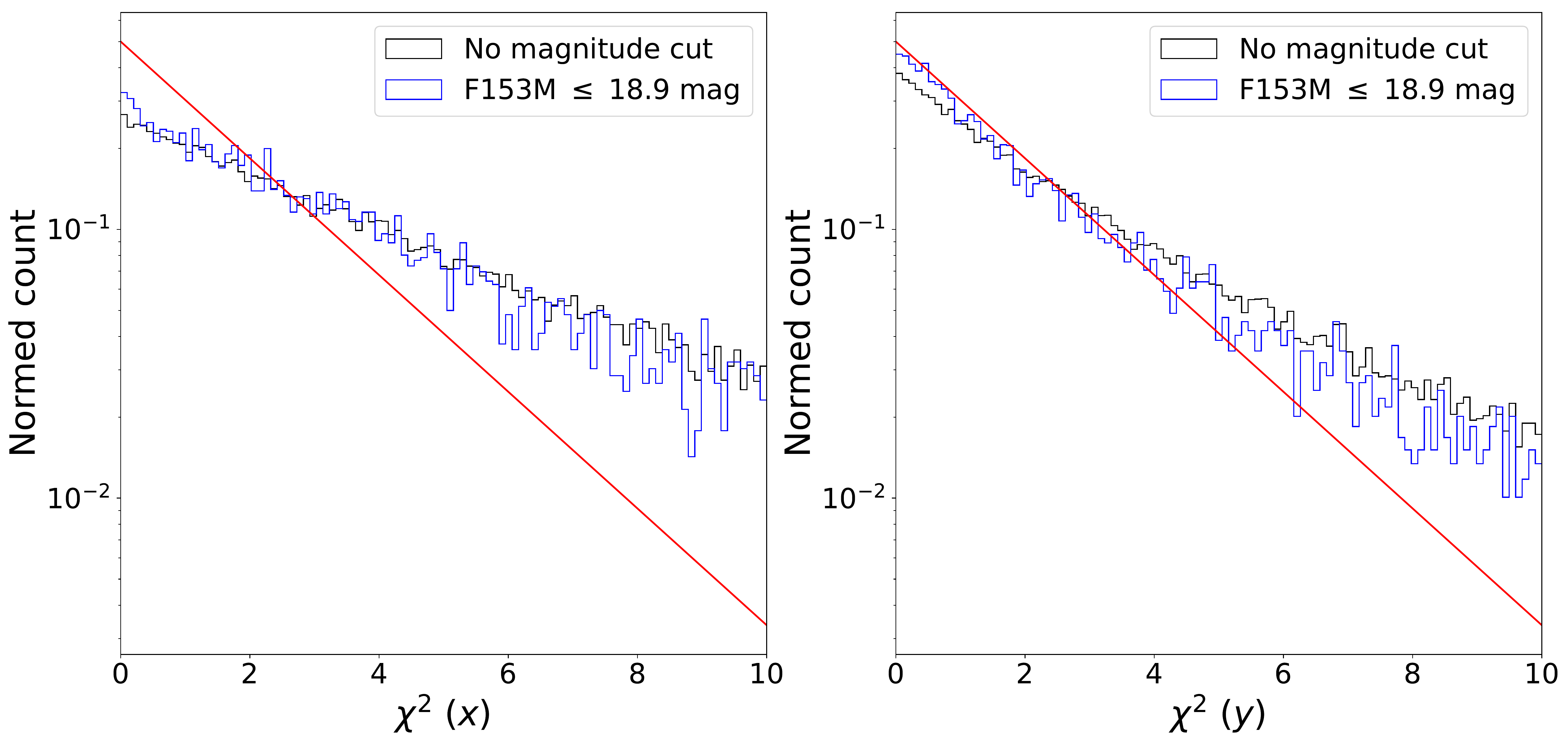}
\centering
\caption{Normed distributions of $\chi^2$ values for $x$ ({\em left}) and $y$ ({\em right}) proper motions, shown both for all 4-epoch detection stars (black) and stars with F153M $\leq 18.9$ mag, the magnitude cut used for profile fitting (blue). Both directions suffer some degree of deviation from the expected $\chi^2$ distribution for high $\chi^2$ values, which is indicative of crowding effects. The distribution of $\chi^2$ values in the $y$ direction is better behaved than in the $x$ direction, where there is an indication that our errors are being somewhat underestimated.}
\end{figure*}

\begin{figure*} \label{elson_full}
\includegraphics[scale=0.43]{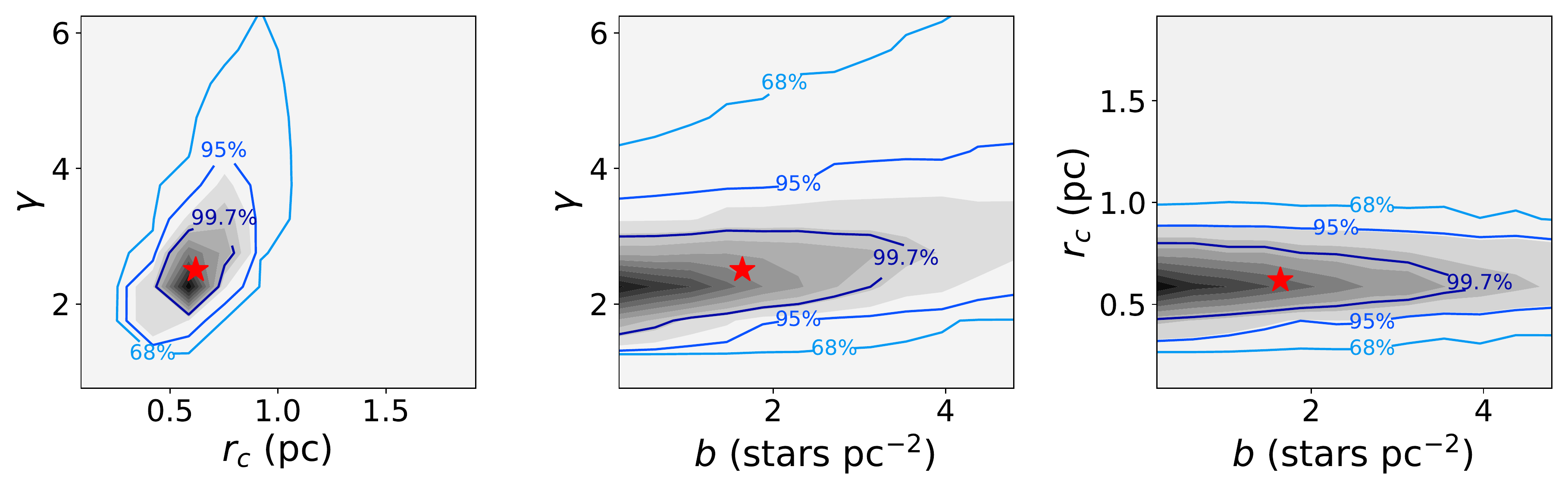}
\centering
\caption{Joint posteriors for the EFF fit to the full cluster, stars with a projected distance of less than 3.2 pc to the center of the cluster and F153M\textsubscript{D} magnitude brighter than 18.9 mag. The \rbold{68\%, 95\%, and 99.7\% confidence contours are shown in blue, and the best-fit parameters is denoted by a red star.}}
\end{figure*}

\begin{figure*} \label{king_full}
\includegraphics[scale=0.43]{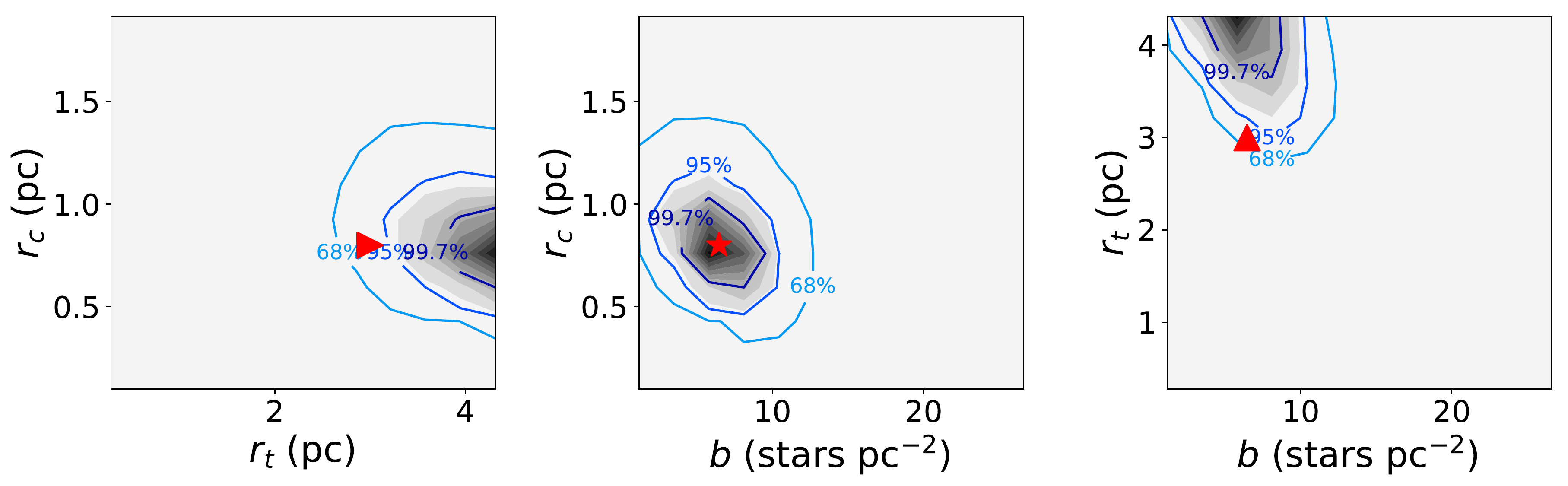}
\centering
\caption{Joint posteriors for the King fit, with \rbold{68\%, 95\%, and 99.7\% confidence contours} shown in blue. \rbold{Best-fit parameters are denoted by red stars, and lower bounds are denoted by red triangles.} Notably, there is no upper constraint on the value of the tidal radius $r_t$ from the data. However, we rule out at $3\sigma$ the possibility of a tidal radius less than or equal to \rbold{3} pc, suggesting that a clear tidal truncation is not present.}
\end{figure*}

\begin{figure*} \label{masssegpost}
\includegraphics[scale=0.43]{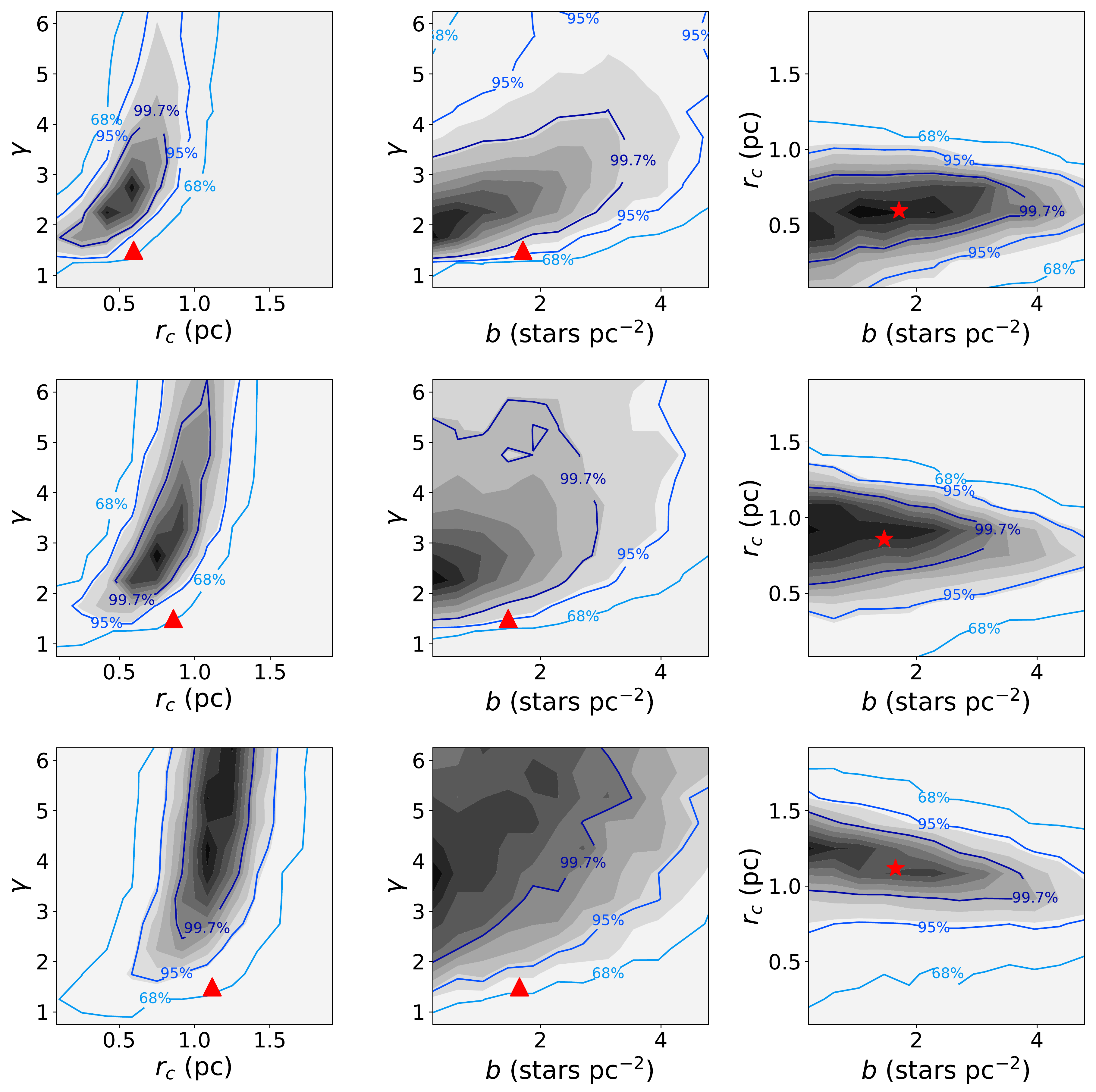}
\centering
\caption{Bivariate posteriors for the EFF profile fit for stars with F153M\textsubscript{D} $<17.8$ (\textit{top}), $17.8$ mag $<$ F153M\textsubscript{D} $<19.1$ mag (\textit{middle}), and $19.1$ mag $<$ F153M\textsubscript{D} $<19.7$ mag (\textit{bottom}). The \rbold{68\%, 95\%, and 99.7\% confidence contours} are shown in blue. \rbold{The best-fit parameters are denoted by red stars, and lower bounds are denoted by red triangles.} \rbold{We} find marginal evidence of mass segregation.}
\end{figure*}

\begin{figure*} \label{directional}
\includegraphics[scale=0.43]{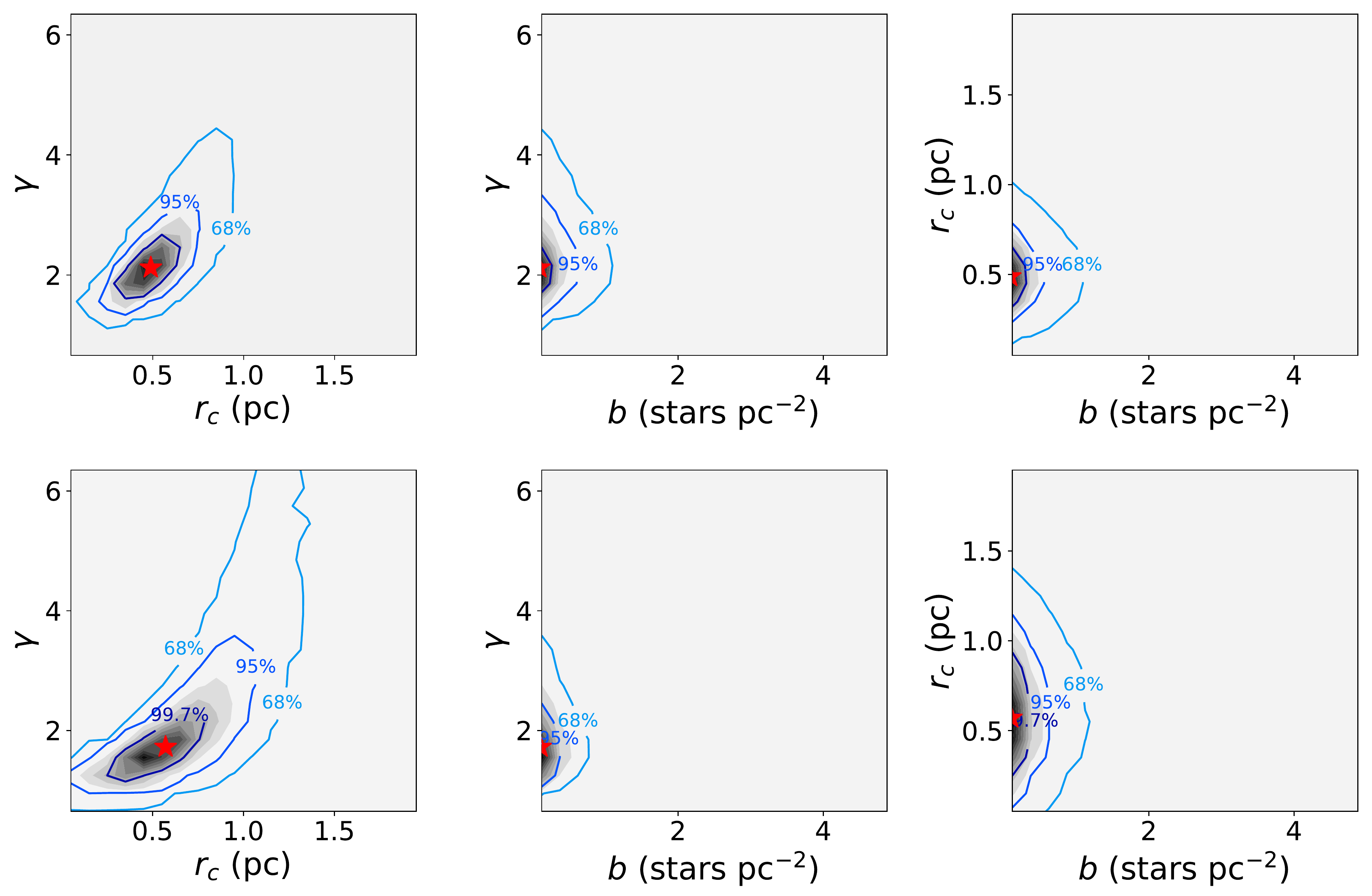}
\centering
\caption{Joint posteriors for the EFF profile fits split into Quintuplet stars parallel to (\textit{top}) and perpendicular to (\textit{bottom}) the direction of the Quintuplet's motion. The \rbold{68\%, 95\%, and 99.7\% confidence contours confidence contours are shown in blue, and the best-fit parameters are denoted by a red star.} Stars are partitioned by two perpendicular lines passing through the center of the cluster lying at 45\degree angles from the Quintuplet's 2D velocity vector.}
\end{figure*}

In this appendix, we present a few summary plots for stars for which we have proper motions. Figure \ref{compareerrors} shows the astrometric errors as well as the F153M and F127M magnitude errors as a function of magnitude. Figure \ref{chisq} shows the $\chi^2$ histogram for the proper motion fit in each direction. We also present the posterior distributions for the radial density profile fits performed in this work. Figure \ref{elson_full} shows the joint posteriors of the EFF model fit for stars lying within 3.2 pc ($82.5''$) with F153M\textsubscript{D} magnitude brighter than 18.7 mag. Figure \ref{king_full} shows the joint posteriors of the King model fit of the same sample. Figure \ref{masssegpost} shows the bivariate posteriors for the mass-binned profile fits in the search for mass segregation. Finally, Figure \ref{directional} shows the joint posteriors for profile fits to the parallel and perpendicular subsamples of the cluster.

% How do I justify X-chi square looking a little weirder

\end{document}